# Maximum Feasible Subsystem Algorithms for Recovery of Compressively Sensed Speech

**FERESHTEH FAKHAR FIROUZEH[1] (Member, IEEE), JOHN W. CHINNECK[1], and SREERAMAN RAJAN[1] (Senior, IEEE)**
[1]Department of Systems and Computer Engineering, Carleton University, Ottawa, ON K1S 5B6 Canada
Corresponding author: Fereshteh Fakhar Firouzeh (e-mail: behnazfakharfirouzeh@cmail.carleton.ca)

**ABSTRACT** The goal in signal compression is to reduce the size of the input signal without a significant loss in the quality of the recovered signal. One way to achieve this goal is to apply the principles of compressive sensing, but this has not been particularly successful for real-world signals that are insufficiently sparse, such as speech. We present three new algorithms based on solutions for the maximum feasible subsystem problem (MAX FS) that improve on the state of the art in recovery of compressed speech signals: more highly compressed signals can be successfully recovered with greater quality. The new recovery algorithms deliver sparser solutions when compared with those obtained using traditional compressive sensing recovery algorithms. When tested by recovering compressively sensed speech signals in the TIMIT speech database, the recovered speech has better perceptual quality than speech recovered using traditional compressive sensing recovery algorithms.

## I. INTRODUCTION

A SPARSE solution is one in which most of the variables have the value zero. The few variables that take nonzero values are called the *support*. Sparse solution estimation or sparse recovery is an important part of *Compressive Sensing* (CS) and plays a major role in reconstructing a compressively acquired signal.

Sparse recovery can be cast as an instance of the *Maximum Feasible Subsystem* problem (MAX FS) [1], which is this: given an infeasible set of linear constraints, find the largest cardinality subset that admits a feasible solution. This is the same as the minimum unsatisfied linear relation problem (MIN ULR) of finding the minimum number of constraints in an infeasible linear system such that its complement is feasible [2]. Finding a maximum feasible subsystem has applications in a wide variety of fields, including machine learning [3], misclassification minimization [4], training of neural networks [2], telecommunications [5], computational biology [6]. MAX FS is NP-hard [7]–[9] but effective polynomial time heuristics are available [1].

Finding a sparse solution to an underdetermined system of linear equations is the central problem in compressive sensing signal recovery, and is cast as an instance of MAX FS [10] as follows: given the system $\mathbf{Ax} = \mathbf{y}$, $\mathbf{x} = 0$, find the maximum cardinality subset of $\mathbf{x} = 0$ that permits a feasible solution to the original system. Several other formulations are also possible.

In compressive sensing, a sparse input signal $\mathbf{a}$ of size $n \times 1$ having $S$ nonzeros ($S$-sparse) is compressed by multiplying it by an $m \times n$ *measurement matrix* $\mathbf{\Phi}$, where $m << n$, to yield the compressed signal $\mathbf{y}$ (also called the *measurement vector*) of size $m \times 1$, i.e. $\mathbf{y} = \mathbf{\Phi a}$, where $\mathbf{\Phi}$ is typically a random matrix. Random matrices are considered in compressive sensing as they have the *Restricted Isometry Property* [11] which is required for signal recovery. The compressed signal $\mathbf{y}$ can now be transmitted or stored much more efficiently because of its greatly reduced size.

The goal of the signal recovery process is to recreate the input signal $\mathbf{a}$ given the compressed signal $\mathbf{y}$ and $\mathbf{\Phi}$. This is an underdetermined system that has multiple solutions, but knowing that the input signal is sparse, the recovery process also tries to return a sparse signal. Unfortunately, recovering a sparse solution from an underdetermined system of linear equations is NP-hard [12], but the sparsity of the recovered signal should be close to the sparsity of the input signal so that the "sparse approximation" is almost an exact recovery.

Mathematically, the sparse approximation problem is to find $\mathbf{x} = arg \min_{\mathbf{x}} \|\mathbf{x}\|_0$ subject to $\mathbf{y} = \mathbf{\Phi x}$ where the number of nonzeros in a vector is commonly expressed as the zero "norm" $\|\mathbf{x}\|_0$. Because the recovery is NP-hard, most algorithms instead minimize some other norm $\|\mathbf{x}\|_p = (\sum_{i=1}^{n} |x_i|^p)^{\frac{1}{p}}, \quad p \geq 1$. Baraniuk [13] evaluated sparse recovery based on $\ell_p$ norm minimization at different values of $p$. Not all norms provide sparse recovery reliably. For instance, $\ell_2$-minimization performs poorly.

$\ell_0$ minimization is a difficult nonconvex problem. Donoho and Hue [14], [15] developed a convex optimization approach called *Basis Pursuit* (BP) which minimizes the $\ell_1$ norm of $\mathbf{x}$. Basis Pursuit is effective in returning an $\mathbf{x}$ that matches the input $\mathbf{a}$ when $\mathbf{a}$ is very sparse [15]–[17], that is BP has small *critical sparsity* (the maximum sparsity at which the algorithm returns sparse solutions reliably). Beyond the critical sparsity, the recovered signal will usually have more non-zero elements than the original sparse signal, and hence will lead to a poor approximation.

It has been shown empirically [18] that using $\ell_p$ norm minimization when $p < 1$ requires fewer measurements (i.e. greater compression) than for $p = 1$. Chartrand and Yin proposed the nonconvex *Iterative Reweighted Least Squares* (IRWLS) algorithm [19] and showed that it needs fewer measurements and has a larger critical sparsity. It can correctly recover less sparse input signals than can be recovered by the unregularized versions of other nonconvex algorithms.

A small critical sparsity means that the recovery algorithm needs a longer measurement vector if it is to return the input vector accurately, so the compressed vector must be larger. BP and greedy algorithms such as *Matching Pursuit* (MP) [20] and *Orthogonal Matching Pursuit* (OMP) [21] are relatively fast, but their low critical sparsity means that they may fail to recover the input signal accurately when the compressed signal is not long enough relative to the sparsity of the input signal. They are thus inappropriate for use with more highly compressed signals. Plumbley [22] proposed the greedy technique *Polytope Faces Pursuit* (PFP) to obtain better recovery of compressed signals which are difficult for MP. This technique is based on the geometry of the polar polytope and uses BP to approximate the sparse solution.

The main issues in sparse recovery are: (i) the small critical sparsities of many widely used recovery algorithms and (ii) the quality of the recovered signals. Existing algorithms can recover the input signal exactly with high probability only when the input signal is very sparse and it is not compressed much [23], otherwise the recovered signal is of low quality. In practical applications, a sparse solution is needed even if these conditions are not met [23]. In practice, the input signal sparsity is not known during the recovery phase; it is either estimated or assumed.

Recognizing that MAX FS solution techniques can be used for sparse recovery, Jokar and Pfetsch [23] compared a number of MAX FS solution algorithms with state-of-the-art algorithms such as BP and OMP for sparse recovery of synthetic signals and concluded that Chinneck's linear programming (LP) based MAX FS solution algorithm [24] provided the best results overall. Surprisingly, MAX FS solution methods have not been adopted for sparse recovery in compressive sensing. This motivates our work here to evaluate MAX FS solution methods for use in the recovery phase of CS for real-world signals.

We investigate the compression of speech signals as they are not sparse by nature [25], and hence are challenging for CS. As a main contribution, we demonstrate that MAX FS-based solution algorithms are able to accurately recover more highly compressed speech signals with better quality, though they require more computation. This is less of an issue in recent years due to the easy availability of computational resources, e.g. via cloud computing.

Our experiments show that the critical sparsities for BP, OMP, PFP, MP and IRWLS require measurement vectors of length $m > 3.2S$, $2.8S$, $3.2S$, $6.4S$, and $8.5S$ respectively. In contrast, the MAX FS solution algorithms require only $m > 2S$ for accurate recovery of low pass speech segments, a reduction of $37.5\%$, $28.6\%$, $37.5\%$, $68.7\%$ and $76.5\%$ in the length of the compressed signal with respect to BP, OMP, PFP, MP, and IRWLS. They require $m > 2.6S$ for accurate recovery of high pass speech segments, still better than the existing algorithms. We also observe higher quality in the recovered signals. The MAX FS-based sparse recovery algorithms perform well in finding both the positions and the values of the nonzeros. We believe that it is time to consider MAX FS-based solutions for CS recovery.

The remainder of the paper is organized as follows. Section II gives a brief overview of CS and existing CS sparse recovery algorithms, as well as background about MAX FS. New MAX FS solution algorithms for sparse recovery are developed in Section III. The CS-based process for speech signals is provided in Section IV. Experimental setup and empirical results are presented in Sections V and VI. Section VII concludes the paper and outlines our future work.

## II. BACKGROUND

### 1) Signal Acquisition and Sparsification

CS compression requires that the input signal be sufficiently sparse. When it is not sparse, the input signal can be sparsified by applying a suitable basis to produce an $S$-sparse signal $\mathbf{a}$. Many real-world signals can be sparsified by applying the DCT (Discrete Cosine Transform) or DWT (Discrete Wavelet Transform) in which the basis coefficient weights satisfy a power law decay. More precisely, if the given input in the time domain, $\mathbf{f}$, is sparsified using the basis $\mathbf{\Psi}$ as $\mathbf{f}_{n\times 1} = \mathbf{\Psi}_{n\times n}\mathbf{a}_{n\times 1}$ and the coefficients are sorted in descending order such that $|a_1| \geq |a_2| \geq ... \geq |a_n|$ , then the signal is compressible if it satisfies

$$|a_i| \leq Const(i^{-q}) \tag{1}$$

where $Const$ is a constant and $q > 0$. To obtain a $S$-sparse signal, all but the $S$ largest coefficients are set to zero.

Based on [26], no information is lost if the length of the measurement vector $m$ is determined as follows:

$$m \geq C.\mu^2(\mathbf{\Phi}, \mathbf{\Psi}).S.\log n \tag{2}$$

where $C$ is a positive constant and $\mu^2(\mathbf{\Phi}, \mathbf{\Psi})$ is the square of mutual coherence between the measurement and the basis (sparsification) matrices. Mutual coherence ($\mu$) between $\mathbf{\Phi}_{m\times n}$ and $\mathbf{\Psi}_{n\times n}$ is defined as follows [14]:

$$\mu(\mathbf{\Phi}, \mathbf{\Psi}) = \max_{1\leq i\neq j\leq n} | < \phi_i, \psi_j > | \tag{3}$$

where $< .,. >$ denotes the numerical operation of inner product between the column vectors $\phi_i, \psi_j \in \mathbb{R}^n$ of $\mathbf{\Phi}$ and $\mathbf{\Psi}$. Low coherence between the measurement matrix, $\mathbf{\Phi}$ and the sensing matrix, $\mathbf{\Psi}$, leads to a better sparse reconstruction from fewer measurements.

To ensure a good recovery, the number of measurements $m$ is often determined as given below, where $\mu^2(\mathbf{\Phi}, \mathbf{\Psi}) = 1$ [26].

$$m \geq C.S.\log n \tag{4}$$

After determining $m$, the compressed measurement vector, $\mathbf{y}_{m\times 1}$, is obtained by multiplying the signal, $\mathbf{a}_{n\times 1}$, by $\mathbf{\Phi}_{m\times n}$ in the last step of signal acquisition to achieve compression.

#### 2) Sparse Recovery

Sparse recovery algorithms can be broadly classified into three categories: convex relaxations, greedy algorithms, and non-convex optimization techniques [27]. We compare the proposed methods with one example algorithm in each class. BP and IRWLS use convex relaxation and a non-convex optimization technique, respectively, while MP, OMP, and PFP are greedy algorithms. These algorithms are known to provide sparse solutions having good reconstructed signal quality. We review the main steps in these algorithms to clarify their approaches.

##### a: Basis Pursuit (BP)

Chen, Donoho, and Saunders [27] find a sparse vector by minimizing the $\ell_1-$norm:

$$\min ||\mathbf{x}||_1 = \sum_{j=1}^{n} |\mathbf{x}_j| \quad \text{s.t.} \quad \mathbf{\Phi x} = \mathbf{y} \tag{5}$$

This can be converted to a linear program (LP) by a change of variables $x_j = u_j - v_j$, where $u_j$ and $v_j$ are nonnegative:

$$\min \sum_{j=1}^{n} (u_j + v_j) \quad \text{s.t.} \quad \mathbf{\Phi}(\mathbf{u} + \mathbf{v}) = \mathbf{y}, \quad u_j, v_j \geq 0 \tag{6}$$

The resulting LP has $2n$ variables and $m$ equations. Upon solution, each $x_j$ is obtained as $x_j = u_j - v_j$.

##### b: Matching Pursuit (MP)

Matching Pursuit [20] is an iterative greedy algorithm. In each iteration, it selects the column $t$ of $\mathbf{\Phi}$, $\phi_{winner_t}$, that is best aligned with the residual vector, $\mathbf{r}_{t-1}$, where $\mathbf{r}_0 = \mathbf{y}$. $winner_t$ is identified using Eqn. 7 [20].

$$winner_t = arg\ \max_{j=1,\dots,n} |\phi_j^H\ \mathbf{r}_{t-1}| \tag{7}$$

where $(.)^H$ is the hermitian transpose matrix. The support is enlarged by adding the index $winner_t$, $support_t = support_{t-1} \cup winner_t$, and the support matrix $\mathbf{\Phi}_{sup}$ is updated as $\mathbf{\Phi}_{sup_t} = [\mathbf{\Phi}_{sup_{t-1}}\ \phi_{winner_t}]$. If $winner_t \in support_{t-1}$, the support set and the support matrix are unchanged, i.e. $support_t = support_{t-1}$ and $\mathbf{\Phi}_{sup_t} = \mathbf{\Phi}_{sup_{t-1}}$.

After updating the support, the new residual vector and the sparse solution are calculated using Eqn. 8 and Eqn. 9.

$$\mathbf{r}_t = \mathbf{r}_{t-1} - \frac{(\phi_{winner_t}^H \mathbf{r}_{t-1})\phi_{winner_t}}{\|\phi_{winner_t}\|_2^2} \tag{8}$$

$$\widetilde{x}_t(winner_t) = \widetilde{x}_{t-1}(winner_t) + \frac{(\phi_{winner_t}^H \mathbf{r}_{t-1})}{\|\phi_{winner_t}\|_2^2} \tag{9}$$

The algorithm halts when the stopping condition is achieved (e.g. $\|\mathbf{r}_t\| \leq \epsilon$).

##### c: Orthogonal Matching Pursuit (OMP)

Orthogonal Matching Pursuit (OMP) [21] is an improvement of MP. In each iteration, the residual vector $\mathbf{r}_t$ is orthogonal to the columns already selected. Therefore, no columns are selected twice. The inputs to this greedy algorithm are the measurement matrix $\mathbf{\Phi}$ and the measurement vector $\mathbf{y}$ [28]. A new element is selected at each step and $\mathbf{\Phi}_{sup_t}$ has full column rank. The OMP algorithm is summarized as follows;

1) Initialization:
   - Iteration Counter: $t \leftarrow 1$.
   - Residual: $\mathbf{r}_0 \leftarrow \mathbf{y}$.
   - Index set: $support \leftarrow \emptyset$.
   - *support* matrix: $\mathbf{\Phi}_{sup} \leftarrow \emptyset$.
2) Find the index $winner_t$ by solving $winner_t = arg\ \max_{j=1,\dots,n} | < \mathbf{r}_{t-1}, \phi_j > |$.
3) Update the index set, $support_t = support_{t-1} \cup winner_t$, and the *support* matrix $\mathbf{\Phi}_{sup_t} = [\mathbf{\Phi}_{sup_{t-1}}, \phi_{winner_t}]$.
4) Estimate the signal by solving a least-squares problem, $\mathbf{x}_t = arg\ \min_{\mathbf{x}} \|\mathbf{\Phi}_{sup_t}\mathbf{x} - \mathbf{y}\|_2$.
5) Update the measurement vector, $\mathbf{y}_t = \mathbf{\Phi}_{sup_t}\mathbf{x}_t$, and the residual, $\mathbf{r}_t = \mathbf{y} - \mathbf{y}_t$.
6) If $\|\mathbf{r}_t\| > threshold$, increment $t$ and go to Step 2.

**Output**:

- $T$-sparse signal, $a_T$.

The goal is obtaining an output signal having a sparsity $T$ as close as possible to $S$. In OMP, the sparsity of the input signal $S$ can be given to the algorithm as an input. If $S$ is specified, the maximum iteration counter $t$ will be equal to $S$. Otherwise, the algorithm stops when $r_t$ reaches to the defined error tolerance $10^{-5}$.

*d: The Polytope Faces Pursuit (PFP)*

This algorithm [22] performs BP to find the sparse solution of the dual LP $max_{\mathbf{c}}\{\mathbf{y^T c}|\mathbf{\Phi^T c} \leq \mathbf{1}\} \geq 0$. In the style of the MP algorithm it adds one new basis vector at each step. PFP adopts a path following method through the relative interior of the faces of the polar polytope $P^* = \{\mathbf{c}|\mathbf{\Phi}^T\mathbf{c} \leq 1\}$ associated with the dual LP problem and searches for the vertex $\mathbf{c}^* \in P^*$ that maximizes $\mathbf{y}^T\mathbf{c}$. The notation $(.)\dagger$ means pseudo-inverse matrix. The steps of the PFP Algorithm are summarised below [22]:

1) Initialization:
   - Iteration counter: $t \leftarrow 1$.
   - Residual: $\mathbf{r}_0 \leftarrow \mathbf{y}$.
   - Index set: $support \leftarrow \emptyset$.
   - Matrix of *support*: $\mathbf{\Phi}_{sup} \leftarrow \emptyset$.
   - c = 0.
2) Find face:
   $winner_t \leftarrow arg\ \max_{i \notin support_{t-1}}\{(\phi_i^T\mathbf{r}_{t-1})/(1 - \phi_i^T\mathbf{c}_{t-1})|\phi_i^T\mathbf{r}_{t-1} > 0\}$
3) Add constraint:
   - $support_t = support_{t-1} \cup winner_t$.
   - $\mathbf{\Phi}_{sup_t} = [\mathbf{\Phi}_{sup_{t-1}}, \phi_{winner_t}]$.
   - $\mathbf{x}_t \leftarrow (\mathbf{\Phi}_{sup_t})^\dagger\mathbf{y}$
4) If $\mathbf{x}_t < 0$
   - Select $j \in support_t$ such that $\mathbf{x}_{t_j} < 0$; remove $\phi_j$ from $\mathbf{\Phi}_{sup_t}$
   - Update:
     $support_t \leftarrow support_t \backslash\{\mathrm{j}\}, \mathbf{x}_t \leftarrow (\mathbf{\Phi}_{sup_t})^\dagger\mathbf{y}$
5) $c_t \leftarrow (\mathbf{\Phi}_{sup_t})^{\dagger T}\mathbf{1}, \mathbf{y}_t \leftarrow \mathbf{\Phi}_{sup_t}\mathbf{x}_t, \mathbf{r}_t \leftarrow \mathbf{y} - \mathbf{y}_t$
6) If termination condition is met (e.g. sparsity or residual) then exit. Else go to Step 2.

**Output**:
- $T$-sparse signal, $a_T$.

The algorithm stops when the size of *support* reaches the maximum sparsity, $S$, (i.e. if specified in the initialization stage, $t = S$) or if $\max_i \phi_i^T\mathbf{r}^{t-1}$ is smaller than the minimum residual condition, $\theta_{min}$.

*e: Iterative Reweighted Least Squares (IRWLS)*

A nonconvex variant of BP [18] has been shown to provide exact recovery with fewer measurements. The $\ell_1$ norm is replaced by the $\ell_p$ norm,

$$\min_{\mathbf{x}} \|\mathbf{x}\|_p^p \quad \text{s.t.} \quad \mathbf{\Phi x} = \mathbf{y} \tag{10}$$

where $0 < p < 1$. $p \geq 1$ was studied before Rao and Kreutz-Delgado [29] considered $p < 1$, replacing the $\ell_p$ cost function in Eqn.10 by a weighted $\ell_2$ norm:

$$\min_{\mathbf{x}} \sum_{i=1}^{n} w_i x_i^2 \quad \text{s.t.} \quad \mathbf{\Phi x} = \mathbf{y} \tag{11}$$

where the objective function is a first order estimate of the $\ell_p$ such that $w_i = |u_i^{(n-1)}|^{p-2}$. Chartrand and Yin [19] proposed a particular regularization strategy that greatly improved the ability of the reweighted least-squares algorithm to recover sparse signals.

In [19], $\mathbf{\Phi}$ is assumed to have the unique representation property (any $m$ columns are linearly independent) [30]. This property leads to a unique solution of $\mathbf{\Phi x} = \mathbf{y}$ having sparsity $\|\mathbf{x}\|_0 = S$. The approach finds weights based on Eqn. 12 for each iteration $t$.

$$w_i = (x_i^2 + \epsilon_t)^{\frac{p}{2}-1} \tag{12}$$

where $\epsilon_t$ is a sequence converging to zero, $\epsilon_t \in (0, 1), 0 \leq p < 2$ and $\mathbf{y} = \mathbf{\Phi x}$. Then, a unique solution of a convex optimization problem Eqn. 11 is obtained in which $\mathbf{x}_t \rightarrow \mathbf{a}$.

## III. MAX FS SOLUTION ALGORITHMS FOR SPARSE RECOVERY

Finding a sparse solution to a linear system can be cast as an instance of MAX FS [10]: find a MAX FS solution for the system $\mathbf{\Phi x} = \mathbf{y}, \mathbf{x} = \mathbf{0}$ where only constraints in the set $\mathbf{x} = \mathbf{0}$ can be removed in order to achieve feasibility. Jokar and Pfetsch [23] used an alternative formulation based on BP (Eqn. 6), as follows. The support is initially empty. At each iteration the $k$ non-support variables having the largest absolute values of $u_i - v_i$ are candidates for inclusion in the support. Each candidate is tested by temporarily setting the objective function values of its associated $u_i$ and $v_i$ to zero and solving the LP: the candidate giving the largest drop in $Z$ is added to the support by permanently zeroing the objective function coefficients of its associated $u_i$ and $v_i$. The process stops when $Z = 0$; the support consists of those variables whose associated $u_i$ and $v_i$ have objective function coefficients of zero.

Jokar and Pfetsch [23] compared Chinneck's algorithm [24] to a number of others for sparse recovery and concluded that it provided the best results overall. Three recent variants of Chinneck's algorithm [31] are used in this paper for CS sparse recovery. The algorithms may return a support having superfluous members. Some can be removed by postprocessing [23] as follows. First, all non-support $x_j$ are set to zero (or removed from the model) in $\mathbf{y} = \mathbf{\Phi x}$. Next, temporarily force each remaining variable to zero in turn; if there is a feasible solution, then that variable is removed from the support.

The values of the support variables are found by solving a final LP. The system $\Phi^{sol}$ containing only the columns of $\mathbf{\Phi}$ corresponding to the support variables is constructed. Then an LP is solved to obtain the values of $u_j$ and $v_j$:

$$min\ Z = \sum_j (u_j + v_j) \quad s.t. \quad \mathbf{\Phi}^{sol}(\mathbf{u} - \mathbf{v}) = \mathbf{y} \tag{13}$$

where $\mathbf{u} \geq 0, \mathbf{v} \geq 0$. The support values are recovered by reversing the change of variables: $x_j = u_j - v_j$.

Three recent variants (Methods C, B, and M) [31] of Chinneck's method are summarized below.

### A. METHOD C

Method C uses explicit elastic variable zeroing constraints $x_j + e_j^+ - e_j^- = 0$, where $e_j^+$ and $e_j^-$ are nonnegative, resulting in the following elastic LP:

$$min\, Z = \sum_j (e_j^+ + e_j^-) \quad s.t.$$

$$\begin{bmatrix} \mathbf{\Phi} & \mathbf{0}_{m\times n} & \mathbf{0}_{m\times n} \\ \mathbf{I} & \mathbf{I} & -\mathbf{I} \end{bmatrix} \begin{bmatrix} \mathbf{x} \\ \mathbf{e}^+ \\ \mathbf{e}^- \end{bmatrix} = \begin{bmatrix} \mathbf{y}_{m\times 1} \\ \mathbf{0}_{n\times 1} \end{bmatrix} \quad (14)$$

$\mathbf{\Phi}$ is $m \times n$ and $\mathbf{I}$ is $n \times n$. The model has $m + n$ constraints in $3n$ variables. The main features of Method C are:

- There are two lists of candidates, one based on the magnitude of the nonzeros, *CandidatesNZ*, and the other based on the sensitivity of the elastic objective function to the variable zeroing constraint, *CandidatesSens*. Both lists are sorted in decreasing order of magnitude and the top *ListLength* candidates from each list are taken.
- Variable $k$ is added to the support set by removing the corresponding elastic variables, $e_k^+$ and $e_k^-$, from the objective function.

Method C is summarized in Fig. 1.

### B. METHOD B

Method B is summarized in Fig. 2. It uses the change of variables LP formulation as in Eqn. (6) and has $m$ constraints in $2n$ variables. It is identical to the Jokar and Pfetsch implementation except for the objective function weights of the support variables. The algorithm follows the general MAX FS algorithm logic with these features:

- Candidate variables $x_j = u_j - v_j$ are those having an objective function coefficient of 1.0 and a magnitude greater than a stated tolerance ($10^{-6}$ is used in the experiments). The length of the list of candidates is controlled by a parameter *ListLength*, typically set to integer value in the range 1 - 7 [24].
- The objective function coefficients of the winning $u_j$, $v_j$ pair are reset to 0.1 instead of 0 [1]. This encourages support variables towards zero, reducing the need for postprocessing.
- At the final solution, only variables that have nonzero values are included in the support set.

### C. METHOD M

Method M combines method B with Basis Pursuit. BP is very efficient if the input vector $\mathbf{a}$ is sufficiently sparse: it returns the correct solution $\mathbf{x}$ after solving a single LP. BP typically returns either a sparse solution $\mathbf{x}$ with $T$-sparsity $<< m$, or it returns $\mathbf{x}$ with a larger sparsity equal to or close to $m$. It is thus easy to recognize when BP has succeeded. M applies the more time-consuming Method B only if BP fails. M assumes BP failure if the $T$-sparsity of the BP solution is greater than $m - 3$, in which case it runs Method B.

**FIGURE 1** Method C

**STEP 0**: $SupportSet \leftarrow \emptyset$
    Set up elastic LP.
**STEP 1**: Solve elastic LP.
    *CandidatesNZ* ← *ListLength* largest magnitude nonzero variables.
    *CandidatesSens* ← *ListLength* variables having value 0 whose zeroing constraints have the largest magnitude sensitivities.
    *CandidateSet* ← *CandidatesNZ* ∪ *CandidatesSens*
**STEP 2**: *WinnerZ* ← ∞.
    **for** each candidate $k$ in *CandidateSet* :
        Set the objective function coefficients of $e_k^+$ and $e_k^-$ to 0.
        Solve elastic LP.
        **if** $Z = 0$ **then**
            Add variable $k$ to *SupportSet*.
            Exit.
        **end if**
        **if** $Z <$ *WinnerZ* **then**
            *Winner* ← $k$.
            *WinnerZ* ← $Z$.
            *NextCandidatesNZ* ← *ListLength* largest magnitude nonzero variables, excluding support variables and $k$.
            *NextCandidatesSens* ← *ListLength* non-support variables having value 0 whose zeroing constraints have the largest magnitude sensitivities.
            *NextCandidateSet* ← *NextCandidatesNZ* ∪ *NextCandidatedatesSens*
        **end if**
        Set the objective function coefficients of $e_k^+$ and $e_k^-$ to 1.
    **end for**
**STEP 3**: Add *Winner* to *SupportSet*.
    Set the objective function coefficients of $e_{winner}^+$ and $e_{winner}^-$ to 0 permanently.
    *CandidateSet* ← *NextCandidateSet*.
    Go to STEP 2.
**OUTPUT**: *SupportSet* is a small number of variables forming a support for the system of equations.

## IV. SPEECH PROCESSING VIA CS AND MAX FS

Speech is a challenging input for CS as it is not typically sparse and any sparsity varies greatly over time [32]. Our process for speech processing using CS with MAX FS sparse approximation has these main steps:

- Signal Acquisition:
    1) $\mathbf{f}$ is the original speech signal in the time domain.
    2) Remove the silent parts of the input.
    3) Segment the signal into frames of length $n$.
    4) For each segment of signal $\mathbf{f}$:

**FIGURE 2** Method B

**STEP 0**: $SupportSet \leftarrow \emptyset$
**STEP 1**: Solve LP.
*CandidateSet* ← *ListLength* largest nonzero $|u_j - v_j|$
**STEP 2**: *WinnerZ* ← $\infty$.
**for** each candidate variable $k$ in *CandidateSet*:
Set the objective function coefficients of $u_k$ and $v_k$ to 0.
Solve the LP.
**if** $Z = 0$ **then**
Add variable $k$ to *SupportSet*.
Exit.
**end if**
**if** $Z <$ *WinnerZ* **then**
*Winner* ← $k$.
*WinnerZ* ← $Z$.
*NextCandidateSet* ← *ListLength* largest nonzero $|u_j - v_j|$ having objective coefficient 1.0
**end if**
Reset the objective function coefficients of $u_k$ and $v_k$ to 1.0.
**end for**
**STEP 3**: Add *Winner* to *SupportSet*.
Fix the coefficients of $u_{winner}$ and $v_{winner}$ to 0.1 in the objective function permanently.
*CandidateSet* ← *NextCandidateSet*.
Go to STEP 2.
**OUTPUT**: *SupportSet* is a small number of variables forming a support for the system of equations.

a) Take DCT transform.
b) Use only the $S$ largest DCT coefficients to generate an $S$-sparse vector $\mathbf{a}$ of length $n$.
c) Calculate the measurement vector $\mathbf{y} = \mathbf{\Phi a}$, where $\mathbf{\Phi}$ is of size $m \times n$.

- Sparse Approximation:
  1) For each segment of signal $\mathbf{f}$:
     a) Apply a MAX FS sparse approximation algorithm to $\mathbf{\Phi x} = \mathbf{y}$ to find a $T$-sparse solution $\mathbf{x}$ as an approximation to $\mathbf{a}$.
- Speech Signal Recovery:
  1) Apply a reverse DCT transform to $\mathbf{x}$ to recover the speech segment in the time domain.
  2) Concatenate all recovered segments to obtain the reconstructed speech signal, $\mathbf{f}$.

The silent portions of a signal contain no useful information, so removing them decreases processing time and increases recovery accuracy. In our experiments, the word transcription information in the dataset is used to identify the silent parts of the input.

Based on [26], by using a proper sparsifying orthonormal basis $\mathbf{\Psi}$, we have $\|\mathbf{f} - \mathbf{f}_S\|_2 = \|\mathbf{a} - \mathbf{a_S}\|_2$ where $\mathbf{f}_S = \Psi \mathbf{a}_S$. When $\mathbf{a}$ is sparse or compressible, $\mathbf{a}$ is well estimated by using $\mathbf{a_S}$ and, consequently, the error $\|\mathbf{f} - \mathbf{f}_S\|_2$ is small, so all except the $S$ largest components of the compressible signal $\mathbf{a}$ can be removed without much loss [26]. Here, to obtain $\mathbf{a_S}$, the DCT coefficients of each segment are sorted in descending order of magnitude; these decay rapidly to zero if the signal is compressible. The $S$ largest coefficients are selected by thresholding. The threshold used here is 1.3 times the mean of all DCT coefficients in a segment and was fixed after examining over 100 different speech segments from the database used in this work.

## V. EXPERIMENTAL SETUP

### A. SPEECH SAMPLES

Examples are drawn from the TIMIT database of speech samples that includes time-aligned orthographic, phonetic and word transcriptions and speech waveforms sampled at 16 $kHz$. [33]. This well-known database has a total of 6300 sentences, 10 sentences spoken by each of 630 speakers, 438 male and 192 female, from 8 major dialect regions of the United States. 96 examples, 48 male and 48 female speakers, are used, covering all 8 dialect regions and all 3 types of sentences. The silent parts of each input speech signal are removed based on the word transcription information in the TIMIT database.

### B. SAMPLING AND MEASUREMENT

The signals are sampled at 16 $kHz$. Speech signals are typically segmented into frames of size $10ms$-$30ms$ due to their non-stationary characteristics. In this paper, speech signals are divided into segments of $16ms$, $n = 256$ with $CompressionRatio(CR) = (1 - \frac{m}{n}) \times 100$ equal to 50%. We study 50% CR.

In the signal acquisition stage, two types of random measurement matrices $\mathbf{\Phi}$ are used to compress the speech signal: *Random Normalized Matrices* (RNM) and *Random Gaussian Matrices* (RGM).

### C. SOFTWARE

All algorithms are implemented in Matlab version 2018, running in a Windows 10 environment. The linear programming solver is MOSEK via the MOSEK Optimization Toolbox for Matlab version 8.1.0.56 [34]. Comparison algorithms were implemented using SparseLab [35], except for IRLWS which uses the code available in [36], [37].

### D. HARDWARE

The computations are carried out on a 3.40 $GHz$ Intel core $i7$ machine with 16.0 $GB$ RAM, running Windows 10.

### E. EXPERIMENTS

Two sets of experiments are conducted. The first set demonstrates that MAX FS has the highest critical sparsity among the algorithms considered. The second set demonstrates that the signals recovered using MAX FS-based algorithms are superior to those recovered by other algorithms.

For the first set of experiments, the speech signals were first divided into two groups: signals that have energy concentration in the "low frequency region" (*low pass*) and signals with energy concentration in the "high frequency region" (*high pass*). A speech signal is low pass if the first 100 DCT coefficients (low frequencies) contribute more to the total energy in the signal than the rest. A speech signal is high pass if the components after the $100^{th}$ coefficient contribute significantly to the total energy of the signal (say 95% of the total energy). 10 *low pass* and 10 *high pass* male and female speech segments were selected for this experiment. Examples of low pass and high pass segments are shown in Fig. 3.

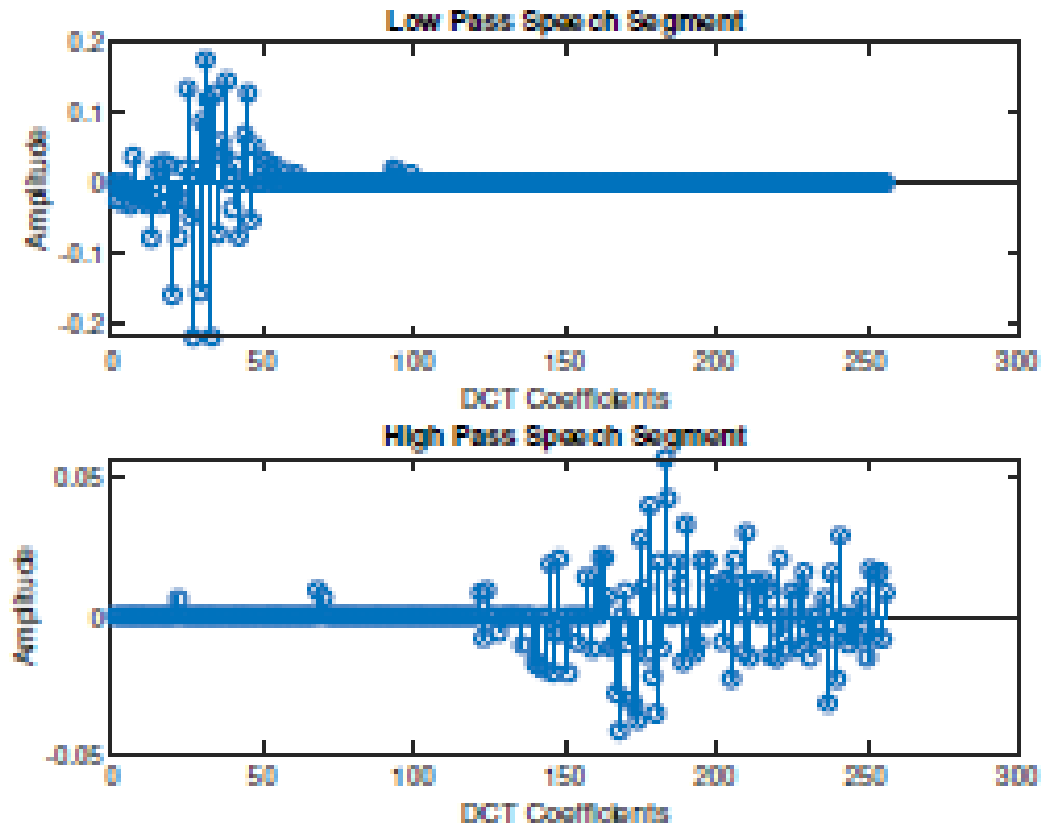

FIGURE 3: Example low and high pass speech signal segments

### F. EVALUATION METRICS

Different evaluation metrics are used for the two sets of experiments.

For the first set of experiments, the recovered signal sparsity is compared with the input signal sparsity. The speech recovery is successful if $T$, the number of nonzeros in the recovered sparse vector, is identical to $S$, the number of nonzeros in the DCT input signal. We record the average $T$-sparsity of the recovered DCT signals over 10 trials, $T_{average}$ at various values of $S$. The number of successful recoveries is recorded. The $GeometricMean$ (GM) of the average $T$-sparsity (Eqn. 15) is used to compare algorithms, following [23].

$$GM = (\prod_{i=1}^{E_{tot}} T_{average_i})^{\frac{1}{E_{tot}}} \tag{15}$$

where $E_{tot}$ is the total number of entries.

The second experiment also evaluates algorithm performance based on the quality of the recovered speech signals as measured by the *Relative Squared Error* (RSE) (Eqn. 16), *Perceptual Evaluation of Speech Quality* (PESQ) [38], spectrograms and spectra.

$$RSE = \frac{\sum_j (\widetilde{f}_j - f_j)^2}{\sum_j (f_j)^2} \tag{16}$$

PESQ is a standardized algorithm recommended by the International Telecommunication Union (ITU) [39] and used to assess the quality of speech [38]. PESQ constructs a loudness spectrum by applying an auditory transform, which is a psychoacoustic model that projects the signals into a representation of perceived loudness in time and frequency [38]. The loudness spectra of the original input signal are then compared with those of the recovered signal to produce a single number in the range 1 (Bad) to 5 (Excellent) corresponding to the prediction of the perceptual mean opinion score.

### G. COMPARATORS

We compare the new MAX FS methods B, C and M with representative algorithms from three different categories of CS sparse recovery algorithms: Basis Pursuit (BP), Matching Pursuit (MP), Orthogonal Matching Pursuit (OMP), Polytope Faces Pursuit (PFP), and Iterative Reweighted Least Squares (IRWLS).

## VI. EXPERIMENTAL RESULTS

### A. CRITICAL SPARSITY OF THE SPARSE RECOVERY ALGORITHMS

A signal recovery algorithm is successful if the recovered signal is exactly the same as the original input signal. Successful recovery becomes harder as the fraction of nonzeros in the input signal increases (i.e. the input is not sparse enough). In our experiments, it is observed that if the output signal $T$-sparsity equals the input signal $S$-sparsity, then the signals are also identical, so we use the matching of the signal sizes as our measure of success. Failures are declared if $T > S$.

The concentration of the DCT coefficients in low and high frequency intervals affects the success of sparse recovery heuristics, so results are analysed for low pass and high pass segments separately.

The results for both RNM and RGM measurement matrices and for low pass and high pass segments are summarized in Table 1 and Table 2. Each cell shows the average output $T$-sparsity $T_{average}$ over 10 segments at given values of input $S$-sparsity, with the number of successes shown in parentheses. The input $S$-sparse signal is constructed by retaining only the $S$ largest DCT coefficients among the 256 input positions. Complete success occurs when $T = S$ in all 10 trials and is indicated in boldface. The last three rows in the tables have the following meanings: "Tot. Succ." shows the total number of successes, "Min $M$" shows the minimum number of measurements required for each algorithm, and "GM" indicates the geometric mean over each column. Algorithms having smaller GMs provide sparser solutions.

Table 1 shows that all algorithms except IRWLS and MP perform very well for $S \leq 35$. MP succeeds completely only when $S \leq 20$ and the measurement matrix is RGM. IRWLS fails for all $S$ for RNM and its critical sparsity is 15 while using RGM. Failures increase with larger $S$, as expected. The three MAX FS algorithms produce the sparsest solutions in geometric mean and fail only when $S > 65$. The general outcome is similar in Table 2, though the algorithms are less successful for the high pass segments. Methods B, M and C again provide better results than the others.

The geometric means from Tables 1 and 2 are summarized in Fig. 4 to compare the effect of RNM vs. RGM. Existing algorithms show better performance on signals compressed using RGM. In contrast, the very best performance is seen for the MAX FS algorithm C when using RNM.

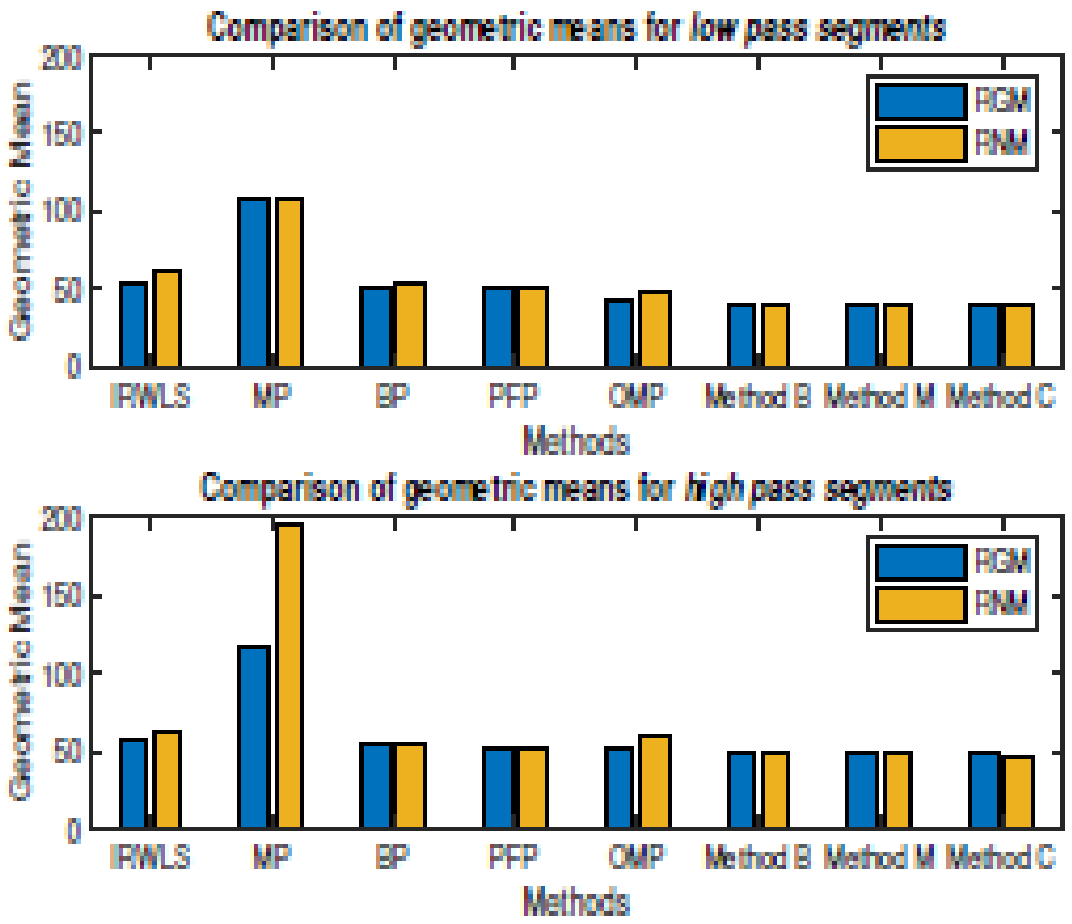

FIGURE 4: Compare the effect of RNM vs. RGM based on geometric means.

Fig. 5 (for RGM) and Fig. 6 (for RNM) summarize the algorithms successes for low pass segments as the input sparsity varies. All algorithms have more failures as $S$-sparsity increases. IRWLS is the worst followed by MP. BP and PFP have roughly the same performance. OMP outperforms all other existing algorithms. The MAX FS recovery methods provide the best results, succeeding in all runs until $S = 65$. Success drops off after $m = 2S$ as expected. Method C provides more successes than Methods B and M for $S = 75$. The MAX FS methods never fail completely even at $S = 80$.

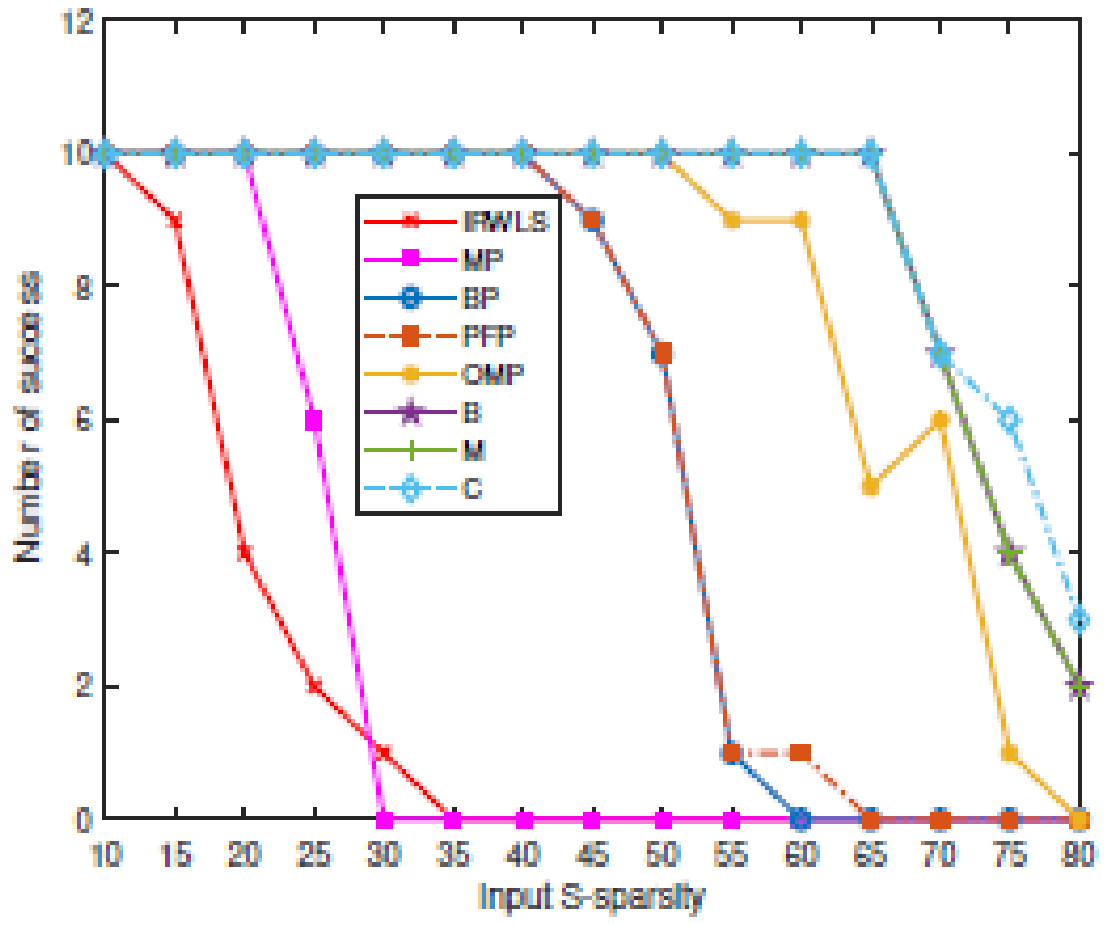

FIGURE 5: Number of successes vs. $S$-sparsity for RGM and low pass segments.

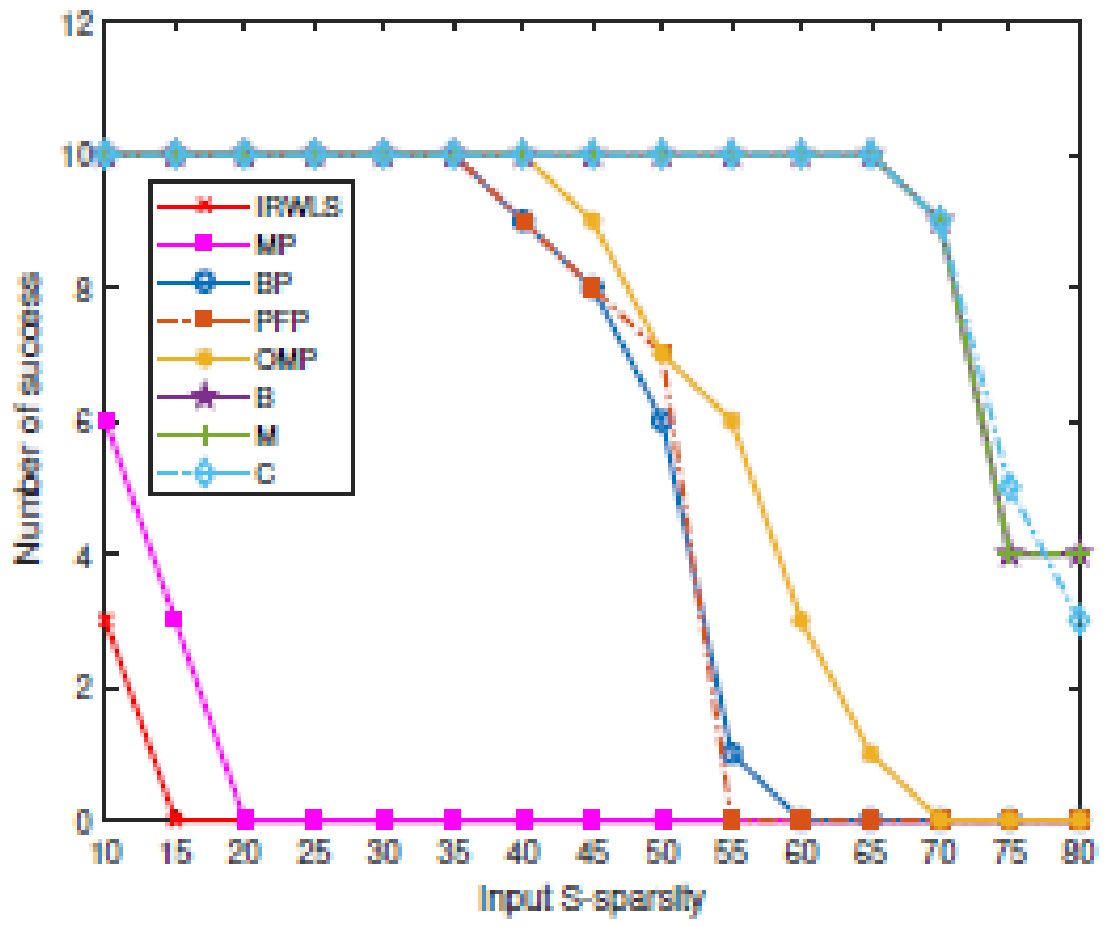

FIGURE 6: Number of successes vs. $S$-sparsity for RNM and low pass segments.

### B. QUALITY OF THE RECOVERED SPEECH SIGNALS

The quality of recovered speech signal depends on the recovery of the sparse DCT coefficients as described previously. 48 male and 48 female speech signals of different lengths are considered. Although RNM provides better results for the MAX FS algorithms, $\Phi$ is RGM since this is preferred by the existing sparse recovery algorithms.

Each speech signal is segmented into frames of length 256. After taking the DCT of each segment, the $S$ largest coefficients are selected by thresholding, where the threshold in each segment is 1.3 times the mean of all of its DCT coefficients. The sparsity of the entire speech signal is the sum of the sparsities of all of its segments. The speech inputs are compressed at CR$= 50\%$ and then recovered. The performances of the algorithms in approximating the input sparsities of the complete speech signals are shown in Fig. 7. The black box shows the sparsity of all 96 uncompressed speech signals. Blue boxes show the estimated sparsities returned by the recovery algorithms. The sparsities are shown as box-and-whisker plots with the median sparsity as the central mark in the box and the $25th$ and $75th$ percentiles as the box boundaries. The whiskers extend to the most extreme sparsities not considered outliers, and the outliers are plotted using the $'+'$ symbol.

The median sparsities are also listed in the inset text. The MAX FS methods have recovered sparsities that are only slightly larger than the input sparsities, and similar ranges. The median recovered sparsities obtained using OMP and the $25th$ percentile of BP are in the upper quartile of the original sparsity level. MP returns the worst result among all algorithms, and its lower extreme of recovered sparsity

TABLE 1: Average Recovered $T$-sparsity for Low Pass Speech Segments at $m = 128$.

| Input S | IRWLS | | MP | | BP | | PFP | | OMP | | B | | M | | C | |
|---|---|---|---|---|---|---|---|---|---|---|---|---|---|---|---|---|
| | RGM | RNM | RGM | RNM | RGM | RNM | RGM | RNM | RGM | RNM | RGM | RNM | RGM | RNM | RGM | RNM |
| 10 | $10^{(10)}$ | $14.5^{(3)}$ | $10^{(10)}$ | $65.3^{(6)}$ | $10^{(10)}$ | $10^{(10)}$ | $10^{(10)}$ | $10^{(10)}$ | $10^{(10)}$ | $10^{(10)}$ | $10^{(10)}$ | $10^{(10)}$ | $10^{(10)}$ | $10^{(10)}$ | $10^{(10)}$ | $10^{(10)}$ |
| 15 | $15^{(10)}$ | $19^{(0)}$ | $15^{(10)}$ | $100.3^{(3)}$ | $15^{(10)}$ | $15^{(10)}$ | $15^{(10)}$ | $15^{(10)}$ | $15^{(10)}$ | $15^{(10)}$ | $15^{(10)}$ | $15^{(10)}$ | $15^{(10)}$ | $15^{(10)}$ | $15^{(10)}$ | $15^{(10)}$ |
| 20 | $21^{(4)}$ | $29.8^{(0)}$ | $20^{(10)}$ | $222.3^{(0)}$ | $20^{(10)}$ | $20^{(10)}$ | $20^{(10)}$ | $20^{(10)}$ | $20^{(10)}$ | $20^{(10)}$ | $20^{(10)}$ | $20^{(10)}$ | $20^{(10)}$ | $20^{(10)}$ | $20^{(10)}$ | $20^{(10)}$ |
| 25 | $27.3^{(2)}$ | $39.4^{(0)}$ | $47.3^{(4)}$ | $222^{(0)}$ | $25^{(10)}$ | $25^{(10)}$ | $25^{(10)}$ | $25^{(10)}$ | $25^{(10)}$ | $25^{(10)}$ | $25^{(10)}$ | $25^{(10)}$ | $25^{(10)}$ | $25^{(10)}$ | $25^{(10)}$ | $25^{(10)}$ |
| 30 | $50.1^{(1)}$ | $50.3^{(0)}$ | $169^{(0)}$ | $227.7^{(0)}$ | $30^{(10)}$ | $30^{(10)}$ | $30^{(10)}$ | $30^{(10)}$ | $30^{(10)}$ | $30^{(10)}$ | $30^{(10)}$ | $30^{(10)}$ | $30^{(10)}$ | $30^{(10)}$ | $30^{(10)}$ | $30^{(10)}$ |
| 35 | $50.1^{(0)}$ | $72^{(0)}$ | $196.4^{(0)}$ | $231.1^{(0)}$ | $35^{(10)}$ | $35^{(10)}$ | $35^{(10)}$ | $35^{(10)}$ | $35^{(10)}$ | $35^{(10)}$ | $35^{(10)}$ | $35^{(10)}$ | $35^{(10)}$ | $35^{(10)}$ | $35^{(10)}$ | $35^{(10)}$ |
| 40 | $70.3^{(0)}$ | $72.4^{(0)}$ | $200^{(0)}$ | $229.7^{(0)}$ | $40^{(10)}$ | $48.8^{(9)}$ | $40^{(10)}$ | $48.1^{(9)}$ | $40^{(10)}$ | $40^{(10)}$ | $40^{(10)}$ | $40^{(10)}$ | $40^{(10)}$ | $40^{(10)}$ | $40^{(10)}$ | $40^{(10)}$ |
| 45 | $75.4^{(0)}$ | $76.6^{(0)}$ | $202.9^{(0)}$ | $229.6^{(0)}$ | $52.7^{(9)}$ | $61.6^{(8)}$ | $52.7^{(9)}$ | $60^{(8)}$ | $45^{(10)}$ | $53^{(9)}$ | $45^{(10)}$ | $45^{(10)}$ | $45^{(10)}$ | $45^{(10)}$ | $45^{(10)}$ | $45^{(10)}$ |
| 50 | $91.7^{(0)}$ | $92.2^{(0)}$ | $199.3^{(0)}$ | $229.4^{(0)}$ | $73.3^{(7)}$ | $81.2^{(6)}$ | $72.7^{(7)}$ | $71.7^{(7)}$ | $79.2^{(7)}$ | $72^{(7)}$ | $50^{(10)}$ | $50^{(10)}$ | $50^{(10)}$ | $50^{(10)}$ | $50^{(10)}$ | $50^{(10)}$ |
| 55 | $99.2^{(0)}$ | $100.1^{(0)}$ | $199.7^{(0)}$ | $231^{(0)}$ | $120.7^{(1)}$ | $127.9^{(0)}$ | $118.4^{(0)}$ | $121.1^{(0)}$ | $88.8^{(6)}$ | $82.8^{(6)}$ | $55^{(10)}$ | $55^{(10)}$ | $55^{(10)}$ | $55^{(10)}$ | $55^{(10)}$ | $55^{(10)}$ |
| 60 | $105.1^{(0)}$ | $102.8^{(0)}$ | $203.4^{(0)}$ | $231.6^{(0)}$ | $121.1^{(1)}$ | $121.2^{(0)}$ | $118.5^{(0)}$ | $121.9^{(0)}$ | $104.4^{(3)}$ | $104.4^{(3)}$ | $60^{(10)}$ | $60^{(10)}$ | $60^{(10)}$ | $60^{(10)}$ | $60^{(10)}$ | $60^{(10)}$ |
| 65 | $101.1^{(0)}$ | $104^{(0)}$ | $206.3^{(0)}$ | $234.3^{(0)}$ | $127.8^{(0)}$ | $127.9^{(0)}$ | $122.8^{(0)}$ | $122.7^{(0)}$ | $116.4^{(1)}$ | $117^{(1)}$ | $65^{(10)}$ | $65^{(10)}$ | $65^{(10)}$ | $65^{(10)}$ | $65^{(10)}$ | $65^{(10)}$ |
| 70 | $103.2^{(0)}$ | $105.1^{(0)}$ | $202.5^{(0)}$ | $230^{(0)}$ | $127.9^{(0)}$ | $127.9^{(0)}$ | $122.2^{(0)}$ | $122.5^{(0)}$ | $120.1^{(0)}$ | $122.6^{(0)}$ | $86.8^{(7)}$ | $75.1^{(9)}$ | $86.8^{(7)}$ | $75.1^{(9)}$ | $85.7^{(7)}$ | $74.9^{(9)}$ |
| 75 | $103.8^{(0)}$ | $104.9^{(0)}$ | $202.4^{(0)}$ | $231.6^{(0)}$ | $128^{(0)}$ | $127.9^{(0)}$ | $122.3^{(0)}$ | $122.4^{(0)}$ | $122.6^{(0)}$ | $122.6^{(0)}$ | $104.7^{(4)}$ | $105.6^{(4)}$ | $104.7^{(4)}$ | $105.6^{(4)}$ | $93.6^{(6)}$ | $98.1^{(5)}$ |
| 80 | $103^{(0)}$ | $112^{(0)}$ | $202.4^{(0)}$ | $231.4^{(0)}$ | $127.9^{(0)}$ | $128^{(0)}$ | $121.2^{(0)}$ | $123^{(0)}$ | $122.9^{(0)}$ | $122.8^{(0)}$ | $116.1^{(2)}$ | $106.7^{(4)}$ | $116.1^{(2)}$ | $106.7^{(4)}$ | $114.3^{(2)}$ | $109.1^{(3)}$ |
| Tot. Succ. | 27 | 3 | 34 | 9 | 88 | 83 | 86 | 84 | 97 | 96 | 133 | **137** | 133 | **137** | 136 | **137** |
| Min M | $8.5S$ | $Fails$ | $6.4S$ | $Fails$ | $3.2S$ | $3.7S$ | $3.2S$ | $3.7S$ | $2.8S$ | $3.2S$ | **2S** | **2S** | **2S** | **2S** | **2S** | **2S** |
| GM | 53.8 | 62.1 | 115.4 | 195.0 | 51.7 | 54.2 | 51.5 | 52.8 | 51.4 | 58.4 | 48.1 | 48.0 | 48.1 | 48.0 | 48.2 | **47.2** |

TABLE 2: Average Recovered $T$-sparsity for High Pass Speech Segments at $m = 128$.

| Input S | IRWLS | | MP | | BP | | PFP | | OMP | | B | | M | | C | |
|---|---|---|---|---|---|---|---|---|---|---|---|---|---|---|---|---|
| | RGM | RNM | RGM | RNM | RGM | RNM | RGM | RNM | RGM | RNM | RGM | RNM | RGM | RNM | RGM | RNM |
| 10 | $10^{(10)}$ | $12.8^{(5)}$ | $10^{(10)}$ | $29.1^{(9)}$ | $10^{(10)}$ | $10^{(10)}$ | $10^{(10)}$ | $10^{(10)}$ | $10^{(10)}$ | $10^{(10)}$ | $10^{(10)}$ | $10^{(10)}$ | $10^{(10)}$ | $10^{(10)}$ | $10^{(10)}$ | $10^{(10)}$ |
| 15 | $15^{(10)}$ | $16.7^{(1)}$ | $15^{(10)}$ | $173.1^{(2)}$ | $15^{(10)}$ | $15^{(10)}$ | $15^{(10)}$ | $15^{(10)}$ | $15^{(10)}$ | $15^{(10)}$ | $15^{(10)}$ | $15^{(10)}$ | $15^{(10)}$ | $15^{(10)}$ | $15^{(10)}$ | $15^{(10)}$ |
| 20 | $20.5^{(5)}$ | $29.6^{(0)}$ | $36.7^{(9)}$ | $215.6^{(0)}$ | $20^{(10)}$ | $20^{(10)}$ | $20^{(10)}$ | $20^{(10)}$ | $20^{(10)}$ | $20^{(10)}$ | $20^{(10)}$ | $20^{(10)}$ | $20^{(10)}$ | $20^{(10)}$ | $20^{(10)}$ | $20^{(10)}$ |
| 25 | $26.1^{(3)}$ | $40.7^{(0)}$ | $95.3^{(3)}$ | $222.4^{(0)}$ | $25^{(10)}$ | $25^{(10)}$ | $25^{(10)}$ | $25^{(10)}$ | $25^{(10)}$ | $25^{(10)}$ | $25^{(10)}$ | $25^{(10)}$ | $25^{(10)}$ | $25^{(10)}$ | $25^{(10)}$ | $25^{(10)}$ |
| 30 | $35.8^{(0)}$ | $49.1^{(0)}$ | $155.5^{(2)}$ | $224.6^{(0)}$ | $30^{(10)}$ | $30^{(10)}$ | $30^{(10)}$ | $30^{(10)}$ | $30^{(10)}$ | $30^{(10)}$ | $30^{(10)}$ | $30^{(10)}$ | $30^{(10)}$ | $30^{(10)}$ | $30^{(10)}$ | $30^{(10)}$ |
| 35 | $45.5^{(0)}$ | $59.8^{(0)}$ | $195^{(0)}$ | $228.3^{(0)}$ | $35^{(10)}$ | $35^{(10)}$ | $35^{(10)}$ | $35^{(10)}$ | $35^{(10)}$ | $44^{(9)}$ | $35^{(10)}$ | $35^{(10)}$ | $35^{(10)}$ | $35^{(10)}$ | $35^{(10)}$ | $35^{(10)}$ |
| 40 | $67.3^{(0)}$ | $71.8^{(0)}$ | $199.3^{(0)}$ | $229.9^{(0)}$ | $40^{(10)}$ | $48.8^{(9)}$ | $40^{(10)}$ | $48.2^{(9)}$ | $40^{(10)}$ | $73.3^{(6)}$ | $40^{(10)}$ | $40^{(10)}$ | $40^{(10)}$ | $40^{(10)}$ | $40^{(10)}$ | $40^{(10)}$ |
| 45 | $81.2^{(0)}$ | $77.3^{(0)}$ | $199^{(0)}$ | $231.5^{(0)}$ | $45^{(10)}$ | $61.6^{(8)}$ | $45^{(10)}$ | $60.3^{(8)}$ | $45^{(10)}$ | $92.1^{(4)}$ | $45^{(10)}$ | $45^{(10)}$ | $45^{(10)}$ | $45^{(10)}$ | $45^{(10)}$ | $45^{(10)}$ |
| 50 | $88.7^{(0)}$ | $83.9^{(0)}$ | $202.6^{(0)}$ | $233^{(0)}$ | $81.2^{(6)}$ | $111.9^{(2)}$ | $87.6^{(5)}$ | $99.1^{(3)}$ | $86.5^{(5)}$ | $124.2^{(0)}$ | $50^{(10)}$ | $50^{(10)}$ | $50^{(10)}$ | $50^{(10)}$ | $50^{(10)}$ | $50^{(10)}$ |
| 55 | $102.9^{(0)}$ | $83.3^{(0)}$ | $200.2^{(0)}$ | $227.7^{(0)}$ | $127.9^{(0)}$ | $113.2^{(2)}$ | $102.8^{(3)}$ | $107.6^{(2)}$ | $124.9^{(0)}$ | $124.9^{(0)}$ | $76^{(9)}$ | $90.1^{(5)}$ | $76^{(9)}$ | $90.1^{(5)}$ | $88.8^{(5)}$ | $82.4^{(6)}$ |
| 60 | $121^{(0)}$ | $100.1^{(0)}$ | $201.8^{(0)}$ | $231.2^{(0)}$ | $127.8^{(0)}$ | $127.7^{(0)}$ | $125.3^{(0)}$ | $121.3^{(0)}$ | $125.5^{(0)}$ | $125.5^{(0)}$ | $126.7^{(0)}$ | $111.6^{(2)}$ | $126.7^{(0)}$ | $111.6^{(2)}$ | $123.2^{(0)}$ | $103.3^{(3)}$ |
| 65 | $122.2^{(0)}$ | $117.3^{(0)}$ | $201^{(0)}$ | $229.4^{(0)}$ | $127.8^{(0)}$ | $127.8^{(0)}$ | $125.3^{(0)}$ | $125.3^{(0)}$ | $125.1^{(0)}$ | $125.1^{(0)}$ | $126.2^{(0)}$ | $118.3^{(1)}$ | $126.2^{(0)}$ | $118.3^{(1)}$ | $122.7^{(0)}$ | $116.6^{(1)}$ |
| 70 | $122^{(0)}$ | $121.8^{(0)}$ | $201.4^{(0)}$ | $228.6^{(0)}$ | $127.9^{(0)}$ | $128^{(0)}$ | $125^{(0)}$ | $123.2^{(0)}$ | $124.6^{(0)}$ | $124.6^{(0)}$ | $125.5^{(0)}$ | $126^{(0)}$ | $125.5^{(0)}$ | $126^{(0)}$ | $123.1^{(0)}$ | $122.3^{(1)}$ |
| 75 | $123^{(0)}$ | $124^{(0)}$ | $203^{(0)}$ | $230.2^{(0)}$ | $127.7^{(0)}$ | $127.8^{(0)}$ | $125.1^{(0)}$ | $121.8^{(0)}$ | $125.1^{(0)}$ | $125.1^{(0)}$ | $125.4^{(0)}$ | $124.2^{(0)}$ | $125.4^{(0)}$ | $124.2^{(0)}$ | $123.0^{(0)}$ | $122.0^{(0)}$ |
| 80 | $123.1^{(0)}$ | $126.5^{(0)}$ | $203.4^{(0)}$ | $230.4^{(0)}$ | $127.9^{(0)}$ | $128^{(0)}$ | $125.4^{(0)}$ | $123.3^{(0)}$ | $126^{(0)}$ | $124.9^{(0)}$ | $126^{(0)}$ | $124.9^{(0)}$ | $126^{(0)}$ | $124.9^{(0)}$ | $122.8^{(0)}$ | $121.8^{(0)}$ |
| Tot. Succ. | 28 | 6 | 34 | 11 | 86 | 81 | 88 | 82 | 85 | 69 | 99 | 98 | 99 | 98 | 95 | **101** |
| Min M | $8.5S$ | $Fails$ | $8.5S$ | $Fails$ | $2.8S$ | $3.7S$ | $2.8S$ | $3.7S$ | $2.8S$ | $4.3S$ | **2.6S** | **2.6S** | **2.6S** | **2.6S** | **2.6S** | **2.6S** |
| GM | 56.2 | 61.8 | 115.4 | 195.0 | 51.7 | 54.2 | 51.5 | 52.8 | 50.63 | 58.44 | 48.1 | 48.0 | 48.1 | 48.0 | 48.2 | 47.2 |

is higher than the upper extreme of the original. The MAX FS algorithms are more successful at recovering the original sparsity of the speech signals at 50% compression than any other algorithm considered. They outperform existing sparse recovery methods in estimating sparsity in real-world speech signals, even when the signal is longer than considered in the previous section.

To recover the complete speech signal, all segments are concatenated after taking the inverse DCT. The 96 recovered signals are evaluated using the Relative Squared Error in Fig. 8. The RSE for the MAX FS methods are very small. They provide higher fidelity recovered signals even though their solutions are sparser than those of the other algorithms.

Fig. 9 evaluates the quality of the returned signals using the Perceptual Evaluation Speech Quality (PESQ). The average PESQ score for recovered female speech signals is better than that for recovered male speech signals. For both male and female speech signals, the MAX FS algorithms outperform the others, providing the highest PESQ score of 4.3 for female speech signals. OMP provides the highest average PESQ score among the traditional recovery algorithms, yet

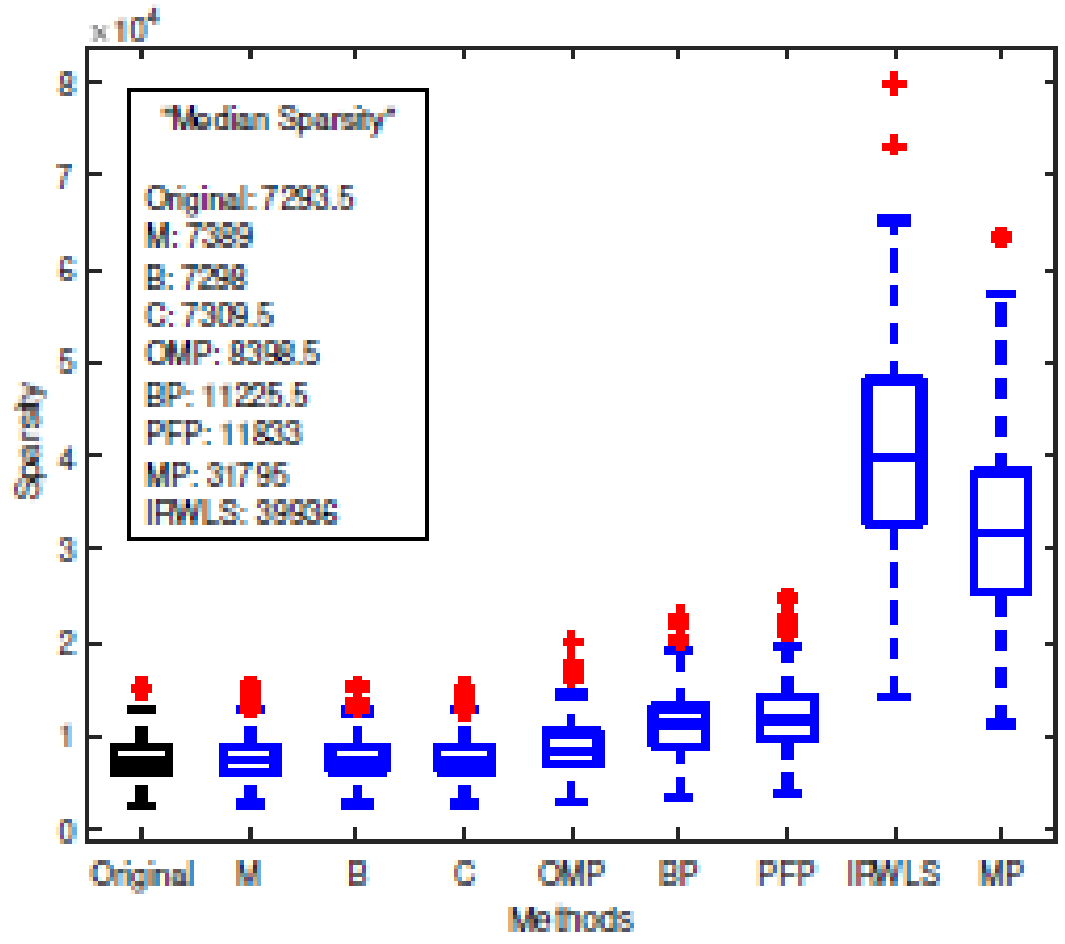

FIGURE 7: Comparison of output $T$-sparsity and input $S$-sparsity for 48 female and 48 male speech signals of differing lengths.

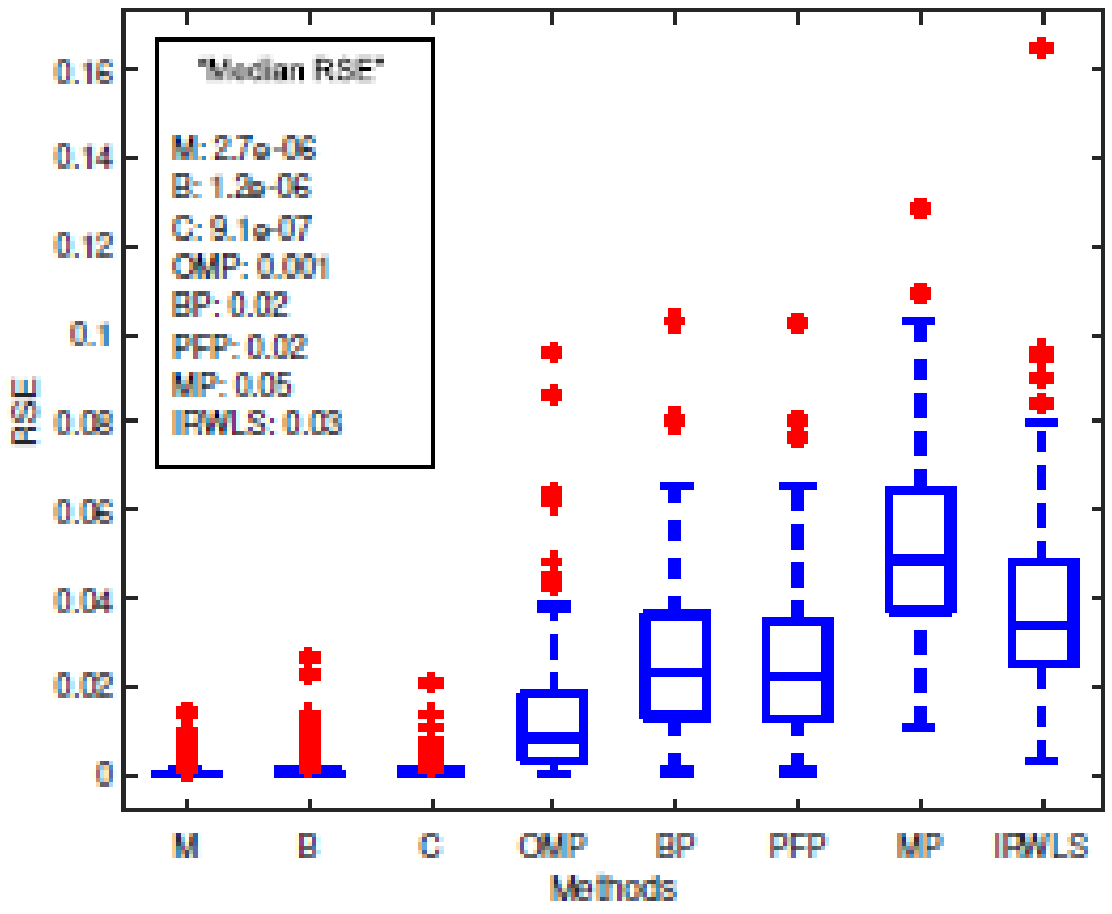

FIGURE 8: Average RSE of 96 recovered signals.

its highest PESQ score is slightly more than 2.5, indicating poor quality.

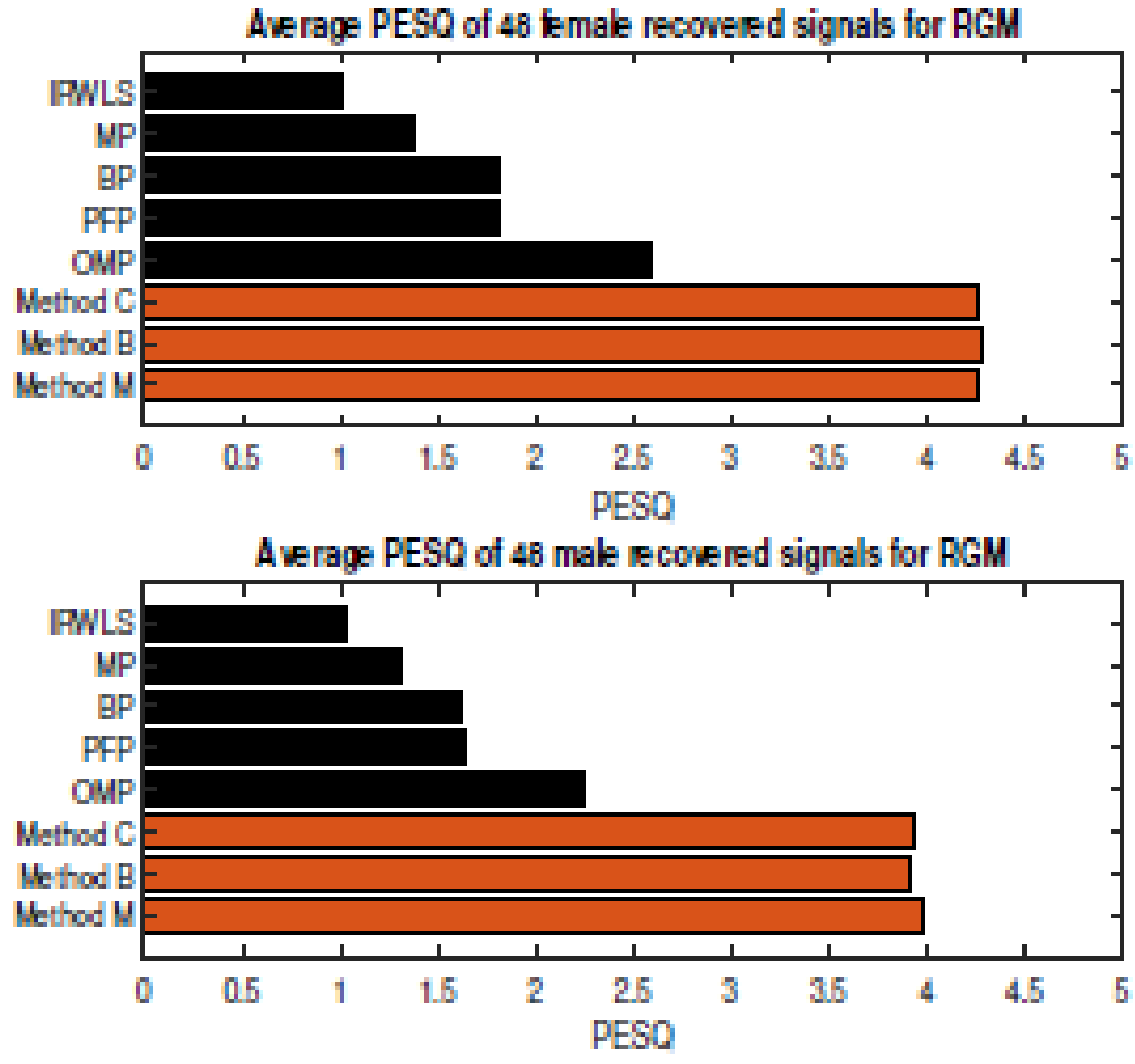

FIGURE 9: Comparing average PESQ for 48 female and 48 male recovered speech signals.

The spectrograms and the frequency responses of the linear predictor coefficients of the recovered and original speech signal of a randomly selected male and randomly selected female speech signal are presented in Figs. 10-12. These figures compare the MAX FS methods with OMP since OMP provided the smallest RSE and sparsity among the existing algorithms as shown in Fig.7 and Fig. 8. The spectrograms of female sample $FDRW0-SA1$ and male sample $MCAL0-SX58$, both the original signal and the recovered signal, are obtained by using Hamming windowing at $16ms$. To improve the FFT performance, a length that is an exact power of two is chosen. The number of data points used for the FFT in each block is 1024.

Fig. 12 shows the good performance of the MAX FS methods in recovering the spectrum of the original signals $FDRW0-SA1$ and $MCAL0-SX58$. The first three formants of the recovered signals follow the first three formants of both female and male original signals. Table 3 compares the recovered sparsity and the formants of the MAX FS methods and OMP with the sparsity and formants of the original female speech signal $FDRW0-SA1$ for CR$=\%50$. The MAX FS methods recover the exact sparsity while following the formants of the original signal. OMP shows good performance in following the original signal formants but fails in estimating sparsity.

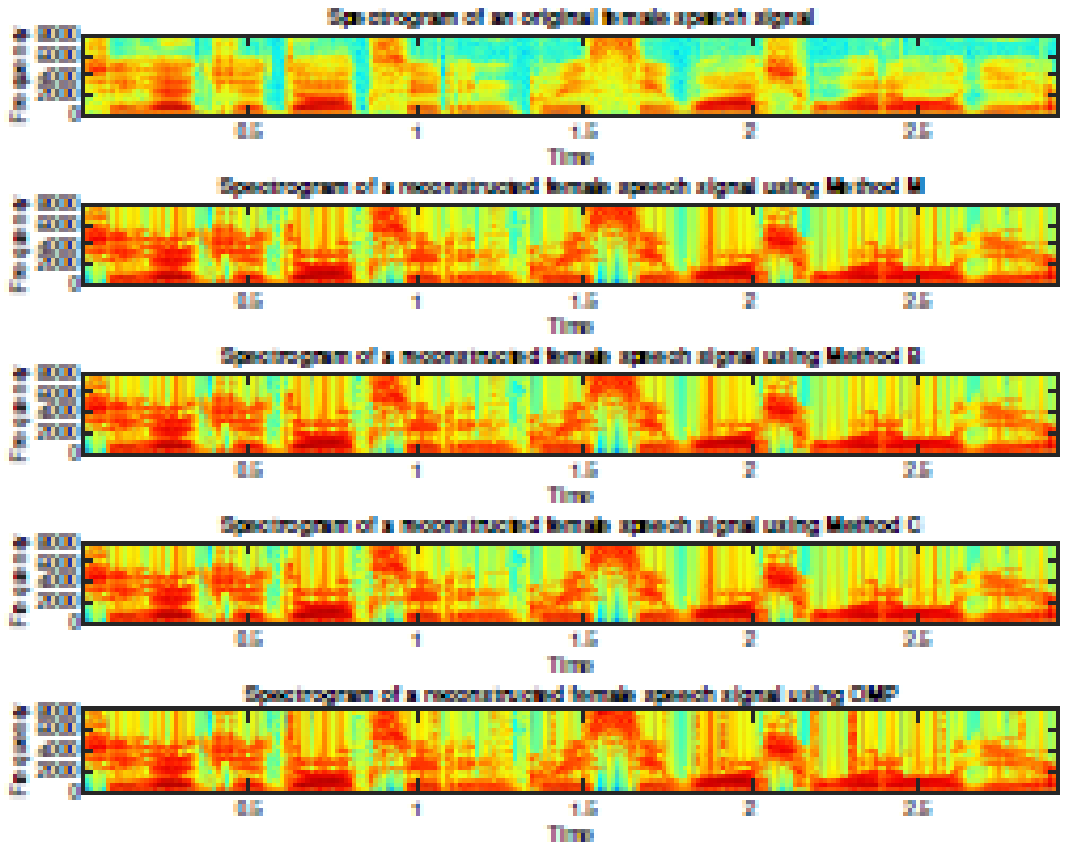

FIGURE 10: Spectrogram of input female speech signal $FDRW0-SA1$ and reconstructed signals.

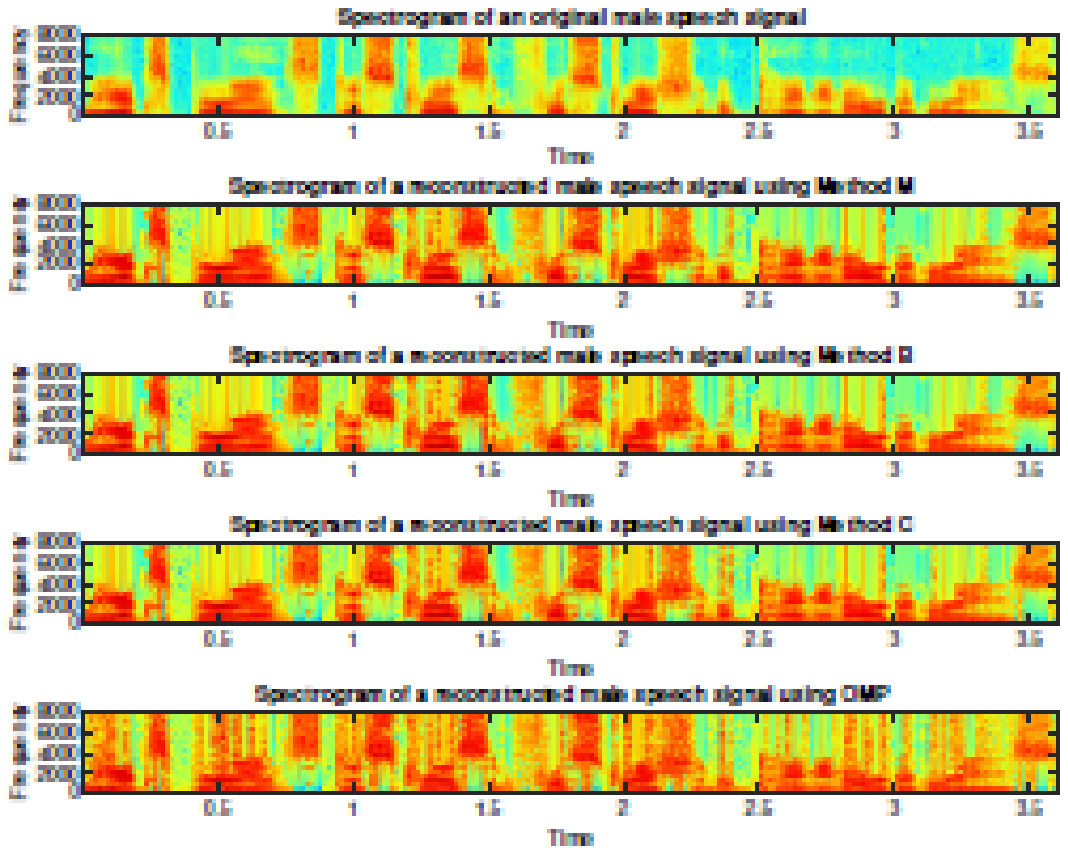

FIGURE 11: Spectrogram of input male speech signal $MCAL0-SX58$ and reconstructed signals.

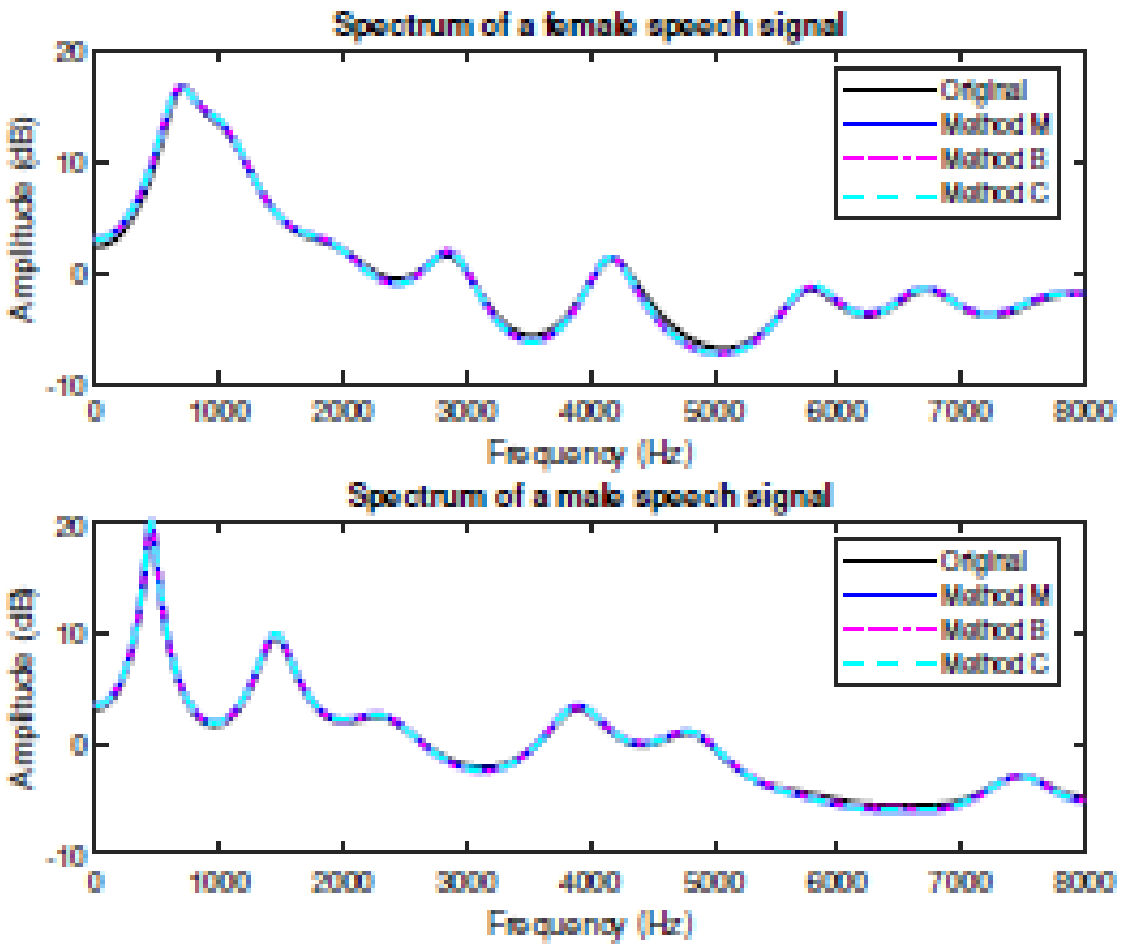

FIGURE 12: Comparing spectrum of female speech signal $FDRW0 - SA1$ and reconstructed signals (Top panel), and spectrum of male speech signal $MCAL0 - SX58$ and reconstructed signals (bottom panel).

TABLE 3: Comparing formants and sparsity level of the original signal with the recovered signal.

| | | Methods | | | |
|---|---|---|---|---|---|
| | Original | B | M | C | OMP |
| Sparsity | 8529 | 8529 | 8529 | 8529 | 9722 |
| F1 | 559.0 | 557.6 | 557.6 | 557.6 | 558.5 |
| F2 | 1012.0 | 1011.5 | 1011.5 | 1011.5 | 1010.2 |
| F3 | 2028.4 | 2016.9 | 2016.9 | 2016.9 | 2019.4 |
| F4 | 2983.0 | 2988.8 | 2988.8 | 2988.8 | 2989.3 |
| F5 | 4218.3 | 4205.8 | 4205.8 | 4205.8 | 4205.8 |

## VII. CONCLUSIONS

This paper describes a technique that uses MAX FS solutions for sparse recovery in compressed sensing speech processing. It shows that MAX FS solution algorithms recover the input signal better than recovery methods commonly used in compressive sensing. MAX FS-based techniques require fewer measurement signals (on the order of $m \geq 2.5S$) for sparse recovery to succeed. Thus when the recovery algorithms are MAX FS-based, higher compression can be used in the measurement phase of compressive sensing.

MAX FS-based recovery requires more computation than most existing recovery algorithms, but its ability to recover more highly compressed signals with higher quality means that it is especially useful for applications such as archiving where it is important to minimize storage size and recovery need not be done in real time. We plan to work towards speeding up the algorithms to give it wider applicability.

We also plan to investigate the application of these new techniques in non-speech applications, e.g. medical uses such as compression and recovery of ECG signals. We are also studying how to adapt the technique to handle noisy signals.

## REFERENCES

[1] J. W. Chinneck, "The maximum feasible subset problem (maxfs) and applications," INFOR: Information Systems and Operational Research, pp. 1–21, 2019.
[2] E. Amaldi, "From finding maximum feasible subsystems of linear systems to feedforward neural network design," Ph.D. dissertation, 1994.
[3] K. P. Bennett and E. J. Bredensteiner, "A parametric optimization method for machine learning," INFORMS Journal on Computing, vol. 9, no. 3, pp. 311–318, 1997.
[4] O. L. Mangasarian, "Misclassification minimization," Journal of Global Optimization, vol. 5, no. 4, pp. 309–323, 1994.
[5] F. Rossi, S. Smriglio, and A. Sassano, "Models and algorithms for terrestrial digital broadcasting," Annals of Operations Research, vol. 107, no. 1-4, pp. 267–283, 2001.
[6] M. Wagner, J. Meller, and R. Elber, "Large-scale linear programming techniques for the design of protein folding potentials," Mathematical programming, vol. 101, no. 2, pp. 301–318, 2004.
[7] N. Chakravarti, "Some results concerning post-infeasibility analysis," European Journal of Operational Research, vol. 73, no. 1, pp. 139–143, 1994.
[8] E. Amaldi and V. Kann, "The complexity and approximability of finding maximum feasible subsystems of linear relations," Theoretical computer science, vol. 147, no. 1-2, pp. 181–210, 1995.
[9] J. K. Sankaran, "A note on resolving infeasibility in linear programs by constraint relaxation," Operations Research Letters, vol. 13, no. 1, pp. 19–20, 1993.
[10] J. W. Chinneck, Feasibility and infeasibility in optimization: algorithms and computational methods. Springer Science+Business Media, 2008.
[11] R. Baraniuk, M. Davenport, R. DeVore, and M. Wakin, "A simple proof of the restricted isometry property for random matrices," Constructive Approximation, vol. 28, no. 3, pp. 253–263, 2008.
[12] B. K. Natarajan, "Sparse approximate solutions to linear systems," SIAM journal on computing, vol. 24, no. 2, pp. 227–234, 1995.
[13] M. Lustig, D. L. Donoho, J. M. Santos, and J. M. Pauly, "Compressed sensing mri," IEEE signal processing magazine, vol. 25, no. 2, p. 72, 2008.
[14] D. L. Donoho and X. Huo, "Uncertainty principles and ideal atomic decomposition," IEEE transactions on information theory, vol. 47, no. 7, pp. 2845–2862, 2001.
[15] E. Candès and T. Tao, "Decoding by linear programming," IEEE Trans. Inf. Theory, vol. 51, no. 12, pp. 4203–4215, 2005.
[16] E. J. Candès, J. K. Romberg, and T. Tao, "Stable signal recovery from incomplete and inaccurate measurements," Communications on Pure and Applied Mathematics: A Journal Issued by the Courant Institute of Mathematical Sciences, vol. 59, no. 8, pp. 1207–1223, 2006.
[17] D. L. Donoho et al., "Compressed sensing," IEEE Transactions on information theory, vol. 52, no. 4, pp. 1289–1306, 2006.
[18] R. Chartrand, "Exact reconstruction of sparse signals via nonconvex minimization," IEEE Signal Processing Letters, vol. 14, no. 10, pp. 707–710, 2007.
[19] R. Chartrand and W. Yin, "Iteratively reweighted algorithms for compressive sensing," in 2008 IEEE International Conference on Acoustics, Speech and Signal Processing. IEEE, 2008, pp. 3869–3872.
[20] S. G. Mallat and Z. Zhang, "Matching pursuits with time-frequency dictionaries," IEEE Transactions on signal processing, vol. 41, no. 12, pp. 3397–3415, 1993.
[21] Y. C. Pati, R. Rezaiifar, and P. S. Krishnaprasad, "Orthogonal matching pursuit: recursive function approximation with applications to wavelet decomposition," in Proceedings of 27th Asilomar conference on signals, systems and computers. IEEE, 1993, pp. 40–44.
[22] M. D. Plumbley, "Recovery of sparse representations by polytope faces pursuit," in International Conference on Independent Component Analysis and Signal Separation. Springer, 2006, pp. 206–213.
[23] S. Jokar and M. E. Pfetsch, "Exact and approximate sparse solutions of underdetermined linear equations," SIAM Journal on Scientific Computing, vol. 31, no. 1, pp. 23–44, 2008.
[24] J. W. Chinneck, "Fast heuristics for the maximum feasible subsystem problem," INFORMS Journal on Computing, vol. 13, no. 3, pp. 210–223, 2001.
[25] U. P. Shukla, N. B. Patel, and A. M. Joshi, "A survey on recent advances in speech compressive sensing," in 2013 International Multi-Conference on Automation, Computing, Communication, Control and Compressed Sensing (iMac4s). IEEE, 2013, pp. 276–280.
[26] E. J. Candès and M. B. Wakin, "An introduction to compressive sampling [a sensing/sampling paradigm that goes against the common knowledge

in data acquisition]," IEEE signal processing magazine, vol. 25, no. 2, pp. 21–30, 2008.
[27] S. S. Chen, D. L. Donoho, and M. A. Saunders, "Atomic decomposition by basis pursuit," SIAM review, vol. 43, no. 1, pp. 129–159, 2001.
[28] J. A. Tropp and A. C. Gilbert, "Signal recovery from random measurements via orthogonal matching pursuit," IEEE Transactions on information theory, vol. 53, no. 12, pp. 4655–4666, 2007.
[29] B. D. Rao and K. Kreutz-Delgado, "An affine scaling methodology for best basis selection," IEEE Transactions on signal processing, vol. 47, no. 1, pp. 187–200, 1999.
[30] I. F. Gorodnitsky and B. D. Rao, "Sparse signal reconstruction from limited data using focuss: A re-weighted minimum norm algorithm," IEEE Transactions on signal processing, vol. 45, no. 3, pp. 600–616, 1997.
[31] J. W. Chinneck, "Sparse solutions of linear systems via maximum feasible subsets," Les Cahiers du GERAD G-2018-104, 2018.
[32] T. V. Sreenivas and W. B. Kleijn, "Compressive sensing for sparsely excited speech signals," in 2009 IEEE International Conference on Acoustics, Speech and Signal Processing. IEEE, 2009, pp. 4125–4128.
[33] J. S. Garofolo, "Timit acoustic phonetic continuous speech corpus," Linguistic Data Consortium, 1993, 1993.
[34] "The mosek optimization software," 2018, last accessed 2019. [Online]. Available: https://www.mosek.com
[35] D. L. Donoho, V. C. Stodden, and Y. Tsaig, "About sparselab," 2007.
[36] C. Chen, J. Huang, L. He, and H. Li, "Preconditioning for accelerated iteratively reweighted least squares in structured sparsity reconstruction," in Proceedings of the IEEE Conference on Computer Vision and Pattern Recognition, 2014, pp. 2713–2720.
[37] ——, "Fast iteratively reweighted least squares algorithms for analysis-based sparsity reconstruction," arXiv preprint arXiv:1411.5057, 2014.
[38] A. W. Rix, J. G. Beerends, M. P. Hollier, and A. P. Hekstra, "Perceptual evaluation of speech quality (pesq)-a new method for speech quality assessment of telephone networks and codecs," in 2001 IEEE International Conference on Acoustics, Speech, and Signal Processing. Proceedings (Cat. No. 01CH37221), vol. 2. IEEE, 2001, pp. 749–752.
[39] I.-T. Recommendation, "Perceptual evaluation of speech quality (pesq): An objective method for end-to-end speech quality assessment of narrow-band telephone networks and speech codecs," Rec. ITU-T P. 862, 2001.

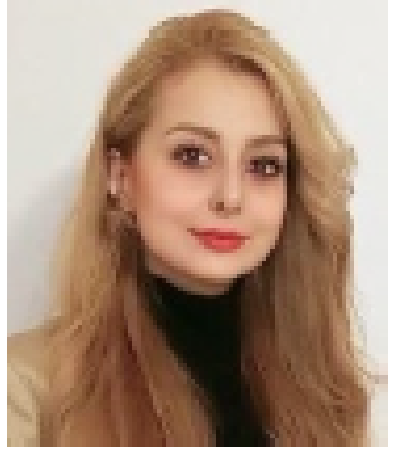

FERESHTEH FAKHAR FIROUZEH is pursuing her PhD in Electrical and Computer Engineering at Carleton University. She has been working on signal processing and compressive sensing (CS) for over 7 years. Her Master's thesis titled "Speech Enhancement Compressive Sensing" was on CS and its application in enhancing and denoising speech signals. During her PhD, she has been developing novel optimization dimensionality reduction techniques based on the Maximum Feasible Subsystem (MAX FS). Her proposed methods can be applied to a variety of areas such as classification, Non-negative Matrix Factorization (NNMF), electrocardiogram signal processing, and speech processing.

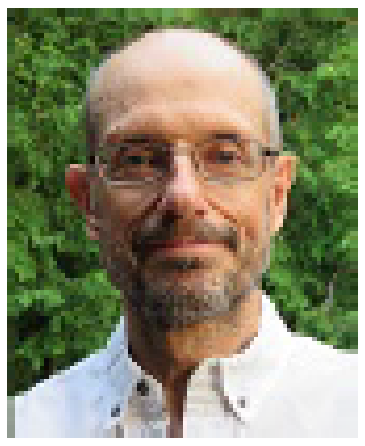

JOHN W. CHINNECK received his PhD in Systems Design Engineering from the University of Waterloo in 1983. He researches the development and application of optimization algorithms. He served as the Editor-in-Chief of *The INFORMS Journal on Computing* from 2007-2012, and as the Chair of the INFORMS Computing Society 2006-2007. He has received Research Achievement Awards from Carleton University, and the Award of Merit from the Canadian Operational Research Society, among other awards.

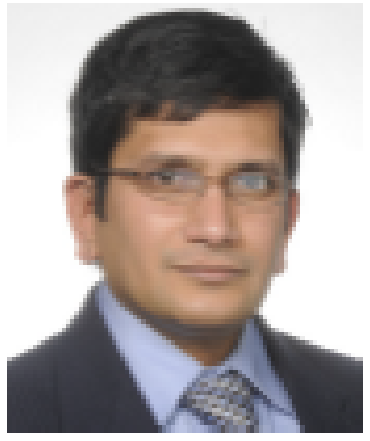

DR. SREERAMAN RAJAN (M'90–SM'06) received the B.E. degree in electronics and communications from Bharathiyar University, Coimbatore, India, in 1987, the M.Sc. degree in electrical engineering from Tulane University, New Orleans, LA, in 1992, and the Ph.D. degree in electrical engineering from the University of New Brunswick, Fredericton, NB, Canada, in 2004.

From 1986 to 1990, he was a Scientific Officer with the Reactor Control Division, Bhabha Atomic Research Center (BARC), Bombay, India, after undergoing an intense training in nuclear science and engineering from its training school. At BARC, he developed systems for control, safety, and regulation of nuclear research and power reactors. During 1997–1998, he carried out research under a grant from Siemens Corporate Research, Princeton, NJ. From 1999 to 2000, he was with JDS Uniphase, Ottawa, ON, Canada, where he worked on optical components and the development of signal processing algorithms for advanced fiber optic modules. From 2000 to 2003, he was with Ceyba Corporation, Ottawa, where he developed channel monitoring, dynamic equalization, and optical power control solutions for advanced ultra-long haul and long haul fiber optic communication systems. In 2004, he was with Biopeak Corporation, where he developed signal processing algorithms for non-invasive medical devices. From December 2004-June 2015, he was a Defense Scientist with the Defence Research and Development Canada, Ottawa, Canada. He joined Carleton University as a Tier 2 Canada Research Chair (Sensors Systems) in its Department of Systems and Computer Engineering in July 2015. He is the Associate Director, Ottawa-Carleton Institute for Biomedical Engineering (OCIBME) since June 2016. He was an Adjunct Professor at the School of Electrical Engineering and Computer Science, University of Ottawa, Ottawa, ON, Canada (July 2010-June 2018) and an Adjunct Professor at the Department of Electrical and Computer Engineering, Royal Military College, Kingston, Ontario, Canada, since July 2015. He is the holder of two patents and two disclosures of invention. He is an author of 160 journal articles and conference papers. His research interests include signal processing, biomedical signal processing, communication, and pattern classification.

He is currently the Chair of the IEEE Ottawa EMBS and AESS Chapters and has served IEEE Canada as its board member (2010-2018). He was awarded the IEEE MGA Achievement Award in 2012 and recognized for his IEEE contributions with Queen Elizabeth II Diamond Jubilee Medal in 2012. IEEE Canada recognized his outstanding service through 2016 W.S. Read Outstanding Service Award. He has been involved in organizing several successful IEEE conferences and has been a reviewer for several IEEE journals and conferences.

• • •